\documentclass[preprint]{aastex631}

\usepackage{enumitem,amssymb}
 \usepackage{amsmath}
\newlist{todolist}{itemize}{2}
\setlist[todolist]{label=$\square$}
\usepackage{pifont}

\usepackage{soul}
\usepackage{colortbl}
\usepackage{multirow}
\graphicspath{{./}{figures/}}

\received{November 11, 2022}
\revised{January 12, 2023}
\accepted{January 18, 2023}
\submitjournal{PSJ}

\shorttitle{Origin and evolution of Mercury's circumsolar dust ring}
\shortauthors{Pokorn\'{y} and et al.}

\begin{document}


\title{Mercury's Circumsolar Dust Ring as an Imprint of a Recent Impact}

\correspondingauthor{Petr Pokorn\'{y}}
\email{petr.pokorny@nasa.gov}

\author[0000-0002-5667-9337]{Petr Pokorn\'{y}}
\affiliation{Department of Physics, The Catholic University of America, Washington, DC 20064, USA}
\affiliation{Astrophysics Science Division, NASA Goddard Space Flight Center, Greenbelt, MD 20771, USA}
\affiliation{Center for Research and Exploration in Space Science and Technology, NASA/GSFC, Greenbelt, MD 20771, USA}

\author[0000-0001-9831-3619]{Ariel N. Deutsch}
\affiliation{Division of Space Science and Astrobiology, NASA Ames Research Center, Mountain View, CA 94035, USA}

\author[0000-0002-2387-5489]{Marc J. Kuchner}
\affiliation{Astrophysics Science Division, NASA Goddard Space Flight Center, Greenbelt, MD 20771, USA}

\begin{abstract}
A circumsolar dust ring has been recently discovered close to the orbit of Mercury. There are currently no hypotheses for the origin of this ring in the literature, so we explore four different origin scenarios here: the dust originated from (1)~the sporadic meteoroid complex that comprises the major portion of the Zodiacal Cloud, (2)~recent asteroidal/cometary activity, (3)~hypothetical dust-generating bodies locked in mean-motion resonances beyond Mercury, and (4)~bodies co-orbiting with Mercury. We find that only scenario~(4) reproduces the observed structure and location of Mercury's dust ring. However, the lifetimes of Mercury's co-orbitals ($<20$~Ma) preclude the primordial origin of the co-orbiting source population due to dynamical instability and meteoroid bombardment, demanding a recent event feeding the observed dust ring. 

We find that an impact on Mercury can eject debris into the co-orbital resonance. We estimate the ages of six candidate impacts that formed craters larger than 40~km in diameter using high-resolution spacecraft data from MESSENGER and find two craters with estimated surface ages younger than 50~Ma. We find that the amount of mass transported from Mercury's surface into the co-orbital resonance from these two impacts is several orders of magnitude smaller than what is needed to explain the magnitude of Mercury's ring inferred from remote sensing. Therefore we suggest that numerous younger, smaller impacts collectively contributed to the origin of the ring. We conclude that the recent impact hypothesis for the origin of Mercury's dust ring is a viable scenario, whose validity can be constrained by future inner solar system missions.
\end{abstract}

\section{Introduction}
\label{SEC:Introduction}
Mercury's neighborhood is the last place in the solar system where one would expect to find a long-term stable circumsolar dust ring. The effects of radiation pressure and Poynting-Robertson drag \citep{Burns_etal_1979} clear all dust and meteoroids smaller than 1 cm inside Mercury's orbit within 1 million years. Collisional lifetimes of centimeter-sized meteoroids in our Zodiacal Cloud (ZC) have been shown to be one-to-two orders of magnitude smaller than the PR drag timescales; however, larger meteoroids are expected to survive much longer \citep{Grun_etal_1985}. Yarkovsky, YORP effect \citep{Bottke_etal_2006}, and thermal cracking \citep{Delbo_etal_2014} affect asteroids of all sizes and severely diminish their survival rates in Mercury's neighborhood. Last but not the least, the presence of Mercury is extremely detrimental for the survival of any object that is crossing Mercury's orbit. Due to Mercury's high  eccentricity, any object not in some orbital resonance with the planet will sooner or later hit it or be expelled from the region between 0.3075 to 0.4667 au by a close encounter with the planet.

For these reasons, it came as a surprise when \citet{Stenborg_etal_2018} reported a circumsolar dust structure near Mercury's orbit found in Solar Terrestrial Relations Observatory (STEREO) data. In their work, \citet{Stenborg_etal_2018} analyzed more than six years of data (December 2007 - March 2014) from the HI-1 instrument onboard the STEREO-A spacecraft. This spacecraft observes the innermost parts of the solar system on solar elongations between $4^\circ$ and $24^\circ$ in white light; this means that the instrument is able to observe the region between heliocentric distances 0.067 au and 0.39 au. After subtracting the background brightness, \citet{Stenborg_etal_2018} discovered a $1.5-2.5\%$ increase in brightness in a band close to the ecliptic and the orbit of Mercury. \citet{Stenborg_etal_2018} attributed this brightness increase to an eccentric circumsolar ring of dust that shows significant variations in the longitudinal direction of the ring. The ring's minimum brightness increase is closely aligned to Mercury's aphelion, whereas the maximum brightness increase is expected to be aligned with Mercury's perihelion. Furthermore, the shape of the circumsolar dust ring in the radial direction seems to be similar to that of Mercury's orbit. Unfortunately, \citet{Stenborg_etal_2018} were not able to map the structure of the entire circumsolar dust ring due to instrumental limitations and the presence of the Galactic plane in the line of sight. 

Mercury's circumsolar dust ring is not the only such structure known in our solar system. A circumsolar dust ring linked to Venus was recently observed \citep{Jones_etal_2013,Jones_etal_2017,Stenborg_etal_2021} and subsequently modelled \citep{Pokorny_Kuchner_2019}. Earth's dust ring has a longer scientific track record \citep{Dermott_etal_1994,Reach_2010}, as well as the main belt dust bands \citep{Dermott_etal_1984,Nesvorny_etal_2006}. Mars, due to its eccentricity and low mass, could potentially host a crescent-shaped dust structure that follows its line of apses \citep{Sommer_etal_2020}. A circumsolar dust ring originating from Jupiter's Trojans is also predicted to exist \citep{Liu_Schmidt_2018,Kuchner_etal_2000}, as well as a wide circumsolar dust ring originating from the Edgeworth-Kuiper belt \citep{Kuchner_Stark_2010,Poppe_2016}.
Observing analogs of Mercury's circumsolar dust ring outside the solar system is currently beyond the capabilities of any astronomical facility due to the rings' faintness and proximity to the host star, as shown via the Atacama Large Millimeter Array (ALMA) observations of Proxima Centauri \citep{Anglada_etal_2017}; however, eccentric structures at greater distances from their host stars in debris disks of various ages are being observed and modelled \citep[e.g.,][]{Faramaz_etal_2014, Pan_etal_2016, Lohne_etal_2017, Pearce_etal_2021}.

Our motivation for our research is rather simple. Currently, there are no hypotheses or theoretical predictions that would explain the origin of Mercury's circumsolar dust ring or any structure that could produce the brightness increase observed by \citet{Stenborg_etal_2018}. We aim to find the most plausible origin hypothesis for Mercury's circumsolar dust ring and lay the foundations for its further exploration.

\section{Observations and constraints of the circumsolar dust ring at Mercury}
\label{SEC:Observations}
\begin{figure}
\epsscale{1.2}
\plotone{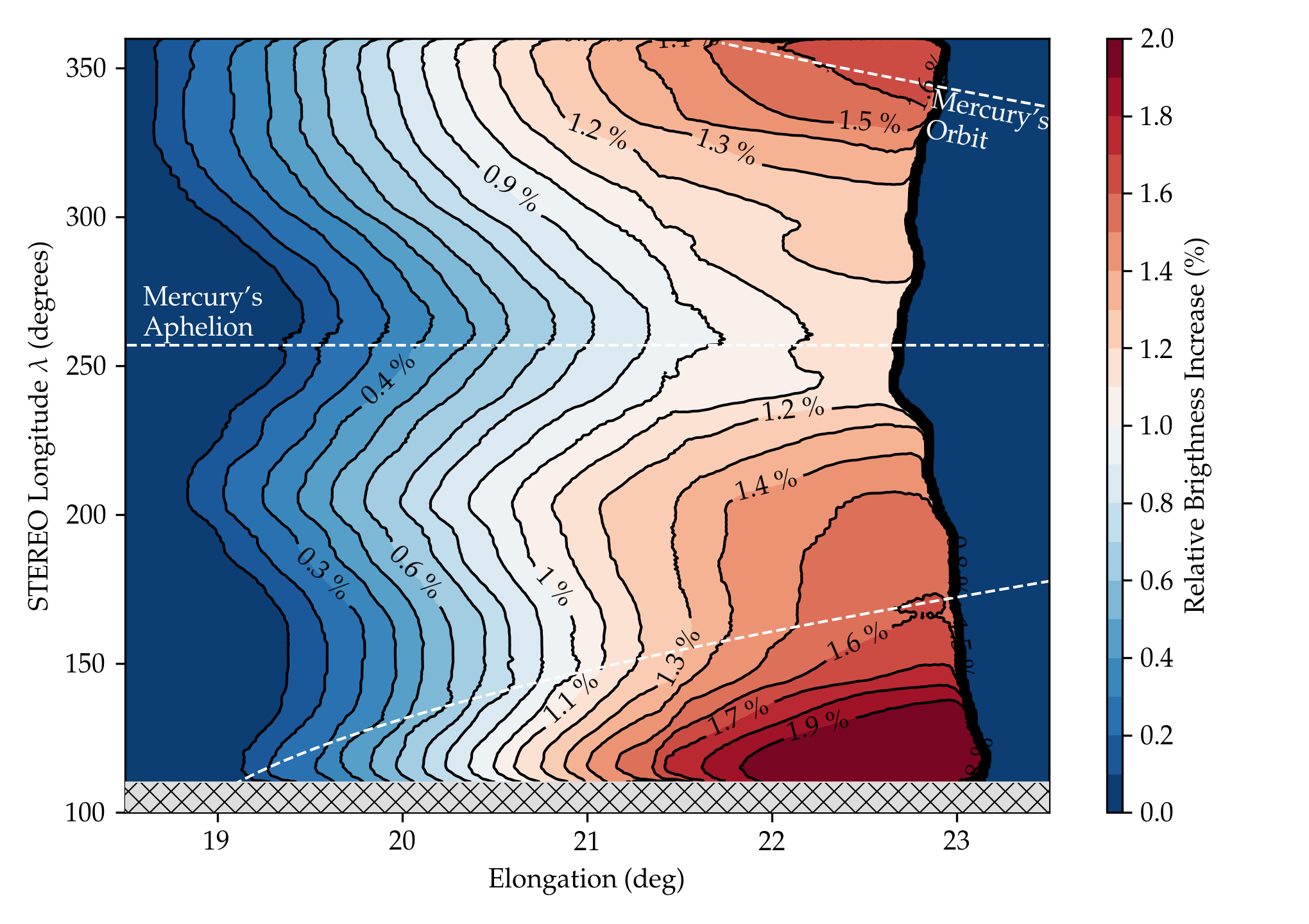}
\caption{Relative brightness increase of the circumsolar dust ring with respect to the brightness of the Zodiacal Cloud. The figure shows the radial (elongation) and azimuthal (STEREO Longitude $\lambda$) distribution of the brightness increase. Mercury's aphelion and orbit locations are shown with white dashed lines. Note that the STEREO observation does not capture the entire structure and is limited to $\sim23^\circ$ in elongation and $>110^\circ$ in STEREO Longitude $\lambda$. Color coding and contours show the asymmetric nature of the dust ring suggesting an eccentric distribution of dust. The data for this plot were adopted from Figure~3 in \citet{Stenborg_etal_2018}. }
\label{FIG:STENBORG_DATA}
\end{figure}

The only currently available observation of Mercury's circumsolar dust ring comes from \citet{Stenborg_etal_2018} and is reproduced in Figure \ref{FIG:STENBORG_DATA}. The other two spacecraft currently able to observe this circumsolar dust ring, Parker Solar Probe \citep{Fox_etal_2016} and Solar Orbiter \citep{Muller_etal_2020}, have yet to provide additional evidence about this peculiar dust structure.

Unfortunately, no latitudinal information about the dust distribution in Mercury's ring is available; we have only information about the relative brightness increase with respect to the solar elongation and the longitude to the STEREO spacecraft, $\lambda$. The STEREO HI-1 instrument that serves as the source of the data in \citet{Stenborg_etal_2018} is observing the eastern side of the Sun at elongations between $4^\circ$ and $24^\circ$. By observing in the eastern direction of the Sun, STEREO can scan the entire ring over a one-year period in small segments. For more discussion, see \citet{Stenborg_etal_2018} and their Appendix.

The structure of Mercury's dust ring brightness profile shows several major features (Fig. \ref{FIG:STENBORG_DATA}). First, it shows an asymmetry in the longitudinal dimension, which points to the eccentric nature of the ring. A circular ring would not have such significant brightness variations in $\lambda$. Furthermore, this longitudinal profile holds for a range of elongations, as shown with contours in Figure \ref{FIG:STENBORG_DATA}. Another major feature is an alignment of the longitudinal portion of the ring with the faintest point on the ring ($\lambda \approx 257^\circ$) with Mercury's aphelion $(\varpi_{AP} =  \Omega_M + \omega_M +180^\circ$ = 257$^\circ$.455). Here, $\varpi$ is the longitude of periapsis, $\Omega$ is the longitude of the ascending node, and $\omega$ is the argument of periapsis. This suggests that the dust particles in the ring are in {some} mean-motion {or apsidal} resonance with Mercury, otherwise, there would be no particular reason for such an alignment \citep{Murray_Dermott_1999_Book}. Thirdly, we see that the brightness profile loosely follows the projected orbit of Mercury that extends beyond the observation elongation limitation ($23^\circ$), and thus we might expect this structure to be more extended than what is shown in Figure \ref{FIG:STENBORG_DATA}. Since the data set is composed of six years of stacked observations, we can conclude that this structure is not significantly affected by the instantaneous position of Mercury.

\subsection{Ring Cross-section and Mass Estimates}
\label{SEC:Ring}
Various observations and models show that $~5-10\%$ of the inner ZC cross-section is contained inside 1 au \citep{Hahn_etal_2002,Nesvorny_etal_2010}. Let us assume that the total cross-section of the ZC is $1-2\times10^{17}$ m$^{2}$ \citep{Nesvorny_etal_2011JFC,Gaidos_1999}, which provides approximately $0.5-2 \times 10^{16}$ m$^{2}$ for the ZC cross-section below heliocentric distance 1~au. The particle density of the inner ZC in the ecliptic $n$, scales with the heliocentric distance $R$ as $n\propto R^{-1.3}$ \citep{Leinert_etal_1981} and we assume the same scaling for the particle cross-section, $\Sigma$. 

Based on \citet{Stenborg_etal_2018}, the observed dust ring is located at the solar elongation region between $\varepsilon = 21^\circ.3 \pm 1^\circ.3$, i.e. between heliocentric distances $r_\mathrm{1}=0.356271$ au and $r_\mathrm{2}=0.400308$ au. Using the total particle cross-section scaling, $\sigma\propto r_\mathrm{hel}^{-1.3}$, we can calculate the ratio between the ZC cross-section inside the boundaries set by Mercury's dust ring and the ZC cross-section interior to 1 au as follows:
\begin{equation}
    \mathcal{R}_\mathrm{ring} = \frac{\int_{r_1}^{r_2} {r_\mathrm{hel}^{-1.3} \mathrm{d} r_\mathrm{hel}}  } {\int_{0.05 \mathrm{au}}^{1 \mathrm{au}} {r_\mathrm{hel}^{-1.3} \mathrm{d} r_\mathrm{hel}}  } = 0.03215,
\end{equation}
where $r_1 = 0.356271$ au, $r_2=0.400308$ au, and we use $r_\mathrm{hel} = 0.05$ au as the boundary of the dust-free zone \citep{Stenborg_etal_2021}. 
Using the average relative brightness increase from \citet{Stenborg_etal_2018} of $\approx 1.5\%$ for Mercury's dust ring, we estimate the dust cross-section of the entire ring to be $\Sigma_\mathrm{ring} = 2.4-9.6 \times 10^{12}$ m$^{2}$; i.e. approximately 1/2000 of the total ZC cross-section inside 1 au.

We can convert the approximate dust cross-section of Mercury's dust ring into a ring mass assuming some size-frequency distribution. Following the equations from  e.g., \citet{Pokorny_etal_2014} or \citet{Pokorny_Kuchner_2019}, we convert between the ring dust cross-section $\Sigma_\mathrm{ring}$ and ring dust mass $M_\mathrm{ring}$:
\begin{equation}
    M_\mathrm{ring} = \frac{6-2\alpha}{12-3\alpha}\rho D_\mathrm{max} \Sigma_\mathrm{ring}\frac{1-\left(\frac{D_\mathrm{min}}{D_\mathrm{max}}\right)^{4-\alpha}}{1-\left(\frac{D_\mathrm{min}}{D_\mathrm{max}}\right)^{3-\alpha}},
\end{equation}
where $\alpha$ is the differential size-frequency distribution index, $\rho$ is the bulk density of dust particles in the ring, $D_\mathrm{min}$ is the smallest particle diameter in the ring, and $D_\mathrm{max}$ is the largest particle diameter. The $D_\mathrm{min}$ and $D_\mathrm{max}$ denote the range where the size-frequency distribution follows a single power-law with an index $\alpha$; i.e. the number of particles larger than $D$ follows $\mathrm{d}N(>D) \propto D^{-\alpha}\mathrm{d}D$. Since we do not have any constraints for any of the parameters, we adopt $\alpha = 3.5$ used for populations in collisional equilibrium \citep{Dohnanyi_1969}, $\rho = 2000$ kg m$^{-3}$, and $D_\mathrm{min}=10~\mu$m $=10^{-5}$~m, and $D_\mathrm{max}=1$ cm $=10^{-2}$ m. This gives us a total mass for Mercury's dust ring of $M_\mathrm{ring} = 1.02 - 4.05 \times 10^{12}$ kg, equivalent to the mass of a single asteroid with a diameter of $D_\mathrm{sphere}= 990-1570$ meters. This is a mass comparable to that of Venus's circumsolar dust ring, as estimated in \citet{Pokorny_Kuchner_2019}. Our estimated value of $M_\mathrm{ring}$ can easily vary by 1-2 orders of magnitude depending on the free parameter setup, such as changing $\alpha$ by unity, or by increasing $D_\mathrm{min}$ to $100~\mu$m.

\section{Origin Scenario Hypotheses}
We now evaluate four different hypotheses for the origin of Mercury's circumsolar dust ring: (A) it occurs naturally as a part of the sporadic meteoroid complex that comprises the major portion of the ZC \citep{Brown_etal_2010,Nesvorny_etal_2010}, (B) it is a product of recent asteroidal and cometary activity, (C) it formed from hypothetical dust-generating populations locked in external mean-motion resonances (MMRs) with Mercury, and (D) it formed from dust and meteoroids created in the 1:1 MMR with Mercury \citep[similar to the circumsolar dust ring co-orbiting with Venus,][]{Pokorny_Kuchner_2019}. 

We assume these four distinctive scenarios represent all potential dust-generating populations in the current solar system. Dust generated from individual asteroids and comets with perihelion distances beyond Mercury's aphelion are captured by Scenarios (A) and (C). This includes dust from abundant dust-generating populations such as Jupiter Trojan and Hilda asteroids, Centaurs \citep{Poppe_2019}, and Kuiper-belt objects \citep{Poppe_2016}. In case that a new abundant population of asteroids on orbits inside Venus' orbit is found \citep{Greenstreet_2020}, such a situation is covered by our Scenario (C).

\section{Methods}
\label{SEC:Methods}
Different origin scenarios analyzed in this paper require different techniques. For Scenario (A), we use the existing models for Jupiter-family comets \citep{Nesvorny_etal_2011JFC}, main belt asteroids \citep{Nesvorny_etal_2010}, Halley-type comets \citep{Pokorny_etal_2014}, and Oort Cloud comets \citep{Nesvorny_etal_2011OCC} that were used to explain various meteoroid phenomena on Mercury \citep{Pokorny_etal_2017_Mercury}, Venus \citep{Janches_etal_2020}, Earth and Moon \citep{Pokorny_etal_2019}, or Ceres \citep{Pokorny_etal_2021}. We also employ models for these populations used in \citet{Pokorny_Kuchner_2019} and \citet{Sommer_etal_2020}.

For Scenarios (B)-(D) we conduct $N$-body simulations using the \texttt{SWIFT} numerical integrator \citep{Levison_Duncan_2013} with the effects of the radiation pressure, Poynting-Robertson drag, and the solar wind included. The solar wind component is simplified to provide a $30\%$ increase to the Poynting-Robertson drag following the results from \citet{Fujiwara_1982}.  As an integration method we use the Regularized Mixed Variable Symplectic (RMVS3) method \citep{Levison_Duncan_1994} with an integration time step of 1 day (86400 seconds), unless stated otherwise. Particles are removed from the simulation once any of the following conditions are fulfilled: a particle's heliocentric distance is less than the solar radius of 0.00468 au, a particle's heliocentric distance is larger than 10,000 au, or a particle hits one of the eight planets. While particles can impact any planet in the simulation we do not include particles' collisions with the ZC \citep{Grun_etal_1985} during the numerical modeling stage to explore the maximum potential of each scenario to create the observed circumsolar dust ring at Mercury.

\subsection{A simple STEREO A HI-1 simulator}
\label{SEC:Simulator}
To reproduce the \citet{Stenborg_etal_2018} observation of Mercury's dust ring, we created a simplified simulator of the HI-1 instrument on-board STEREO A. The level of processing that \citet{Stenborg_etal_2018} implemented is beyond the scope of this article and is not necessary because we can simulate each dust population separately. Moreover, any noise present in any F-corona/ZC model would further decrease the fidelity of our results. Therefore, we try to keep our synthetic telescope model as simple as possible.

We use the spacecraft orbital elements $a=0.9618$ au, $e=0.00583$, $I=0.126^\circ$, $\Omega = 213.8^\circ$, and $\omega=93.8^\circ$ to generate 360 heliocentric position vectors along the orbit uniformly distributed with respect to the spacecraft's ecliptic longitude $\lambda$. From each of these positions, we calculate the relative brightness of each dust particle as:
\begin{equation}
    \mathcal{B} =\frac{\pi D^2}{4} \frac{1}{||\vec{r_\mathrm{par}}-\vec{r_\mathrm{SC}}||^2}\frac{1}{||\vec{r_\mathrm{par}}||^2} \mathcal{P}(\cos\gamma) = \frac{\pi D^2}{4} \frac{1}{||\vec{\Delta}||^2}\frac{1}{||\vec{r_\mathrm{par}}||^2} \mathcal{P}(\cos\gamma),
    \label{EQ:Brightness}
\end{equation}
where $D$ is the particle diameter, $\vec{r_\mathrm{par}}$ is the heliocentric position vector of the particle, $\vec{r_\mathrm{SC}}$ is the heliocentric position vector of the spacecraft, $\mathcal{P}$ is the scattering phase function, $\gamma$ is the Sun-particle-observer angle, $\vec{\Delta} = \vec{r_\mathrm{par}}-\vec{r_\mathrm{SC}}$ is the vector pointing from the spacecraft to the particle, and $||$ denotes the length of the vector. {Note that all particles in our model are much larger than the wavelength range observed by HI-1 instrument; 630-730 nm.}

The scattering phase function $\mathcal{P}$ is calculated following Table 1 from \citet{Hong_1985}, who used a linear combination of three Henyey-Greenstein functions as:
\begin{equation}
    \mathcal{P}(\cos\gamma) = \sum_{i=1}^3\frac{w_i}{4\pi}\frac{1-g_i^2}{\left(1+g_i^2-2g_i\cos\gamma\right)^{3/2}},
\end{equation}
where $w_1=0.665$, $w_2=0.330$, $w_3=0.005$, $g_1=0.70$, $g_2=-0.20$, and $g_3=-0.81$. The scattering angle $\gamma$ is calculated as:
\begin{equation}
    \gamma = \arccos\left(\frac{\vec{\Delta}\cdot \vec{-r_\mathrm{SC}}    }{||\vec{\Delta}|| || \vec{r_\mathrm{SC}}||} \right)
\end{equation}.

Finally, we calculate the particle solar elongation in the ecliptic $\epsilon$ for each particle as the angle between the vernal equinox and the particle's position in the ecliptic:
\begin{equation}
    \epsilon = \arccos\left(\frac{\vec{\Delta_\mathrm{ecl}}\cdot \vec{-r_\mathrm{SC_\mathrm{ecl}}}    }{||\vec{\Delta_\mathrm{ecl}}|| || \vec{r_\mathrm{SC_\mathrm{ecl}}}||} \right)
\end{equation}
where the vector projection to the ecliptic is calculated as $\vec{\Delta_\mathrm{ecl}} = \vec{\Delta}-(\vec{\Delta}\cdot \vec{n_z}) \vec{n_z}$, where $\vec{n_z}=(0,0,1)$ is the normalized vector normal to the ecliptic plane.  This is $\gamma$ projected to the ecliptic. We also need to calculate the direction/sign of the elongation, $\mathcal{D}$, in order to distinguish westward/eastward pointing. The direction, $\mathcal{D}$, is calculated as: 

\begin{equation}
    \mathcal{D} = \mathrm{sign}(\vec{r_{\mathrm{SC}_\mathrm{ecl}\perp}}\cdot \vec{\Delta_\mathrm{ecl}}) = \mathrm{sign}(-r_{\mathrm{SC}_y} \Delta_x + r_{\mathrm{SC}_x} \Delta_y).
\end{equation}

\section{Infeasible Origin Scenarios for Mercury's Ring}
In this Section, we discuss the first three hypotheses (A-C), which we conclude cannot explain the existence of the circumsolar dust ring close to Mercury's orbit. We analyze three difference scenarios: a) accumulation of dust and meteoroids close to Mercury's orbit originating from the sporadic meteoroid complex (Section \ref{SEC:SPORADIC}); b) concentration of dust and meteoroids originating in recent asteroidal and cometary activity (Section \ref{SEC:RECENT_ACTIVITY}); and c) dust and meteoroids originating in hypothetical source populations locked in Mercury's mean-motion resonances.

\subsection{Sporadic meteoroid and dust background}
\label{SEC:SPORADIC}
The sporadic meteoroid and dust background is the main component of the ZC \citep{Koschny_etal_2019}. Numerous modeling \citep{Wiegert_etal_2009,Nesvorny_etal_2010} and observational studies \citep{CampbellBrown_2008, Janches_etal_2015,CarrilloSanchez_etal_2020,Rojas_etal_2021} showed that three main components dominate the inner solar system dust and meteoroid budget with diameters between several micrometers and several millimeters. The main belt asteroids supply the ZC component closest to the ecliptic \citep{Nesvorny_etal_2010}, the short-period/Jupiter-family comets (JFCs) make up the majority of the observable ZC with 80-90\% of the total mass and particle cross-section \citep{Nesvorny_etal_2011JFC}, while the long-period comets (LPCs) supply the broad envelope and particles on retrograde orbits \citep{Nesvorny_etal_2011OCC, Pokorny_etal_2014}. The outer solar sources of dust with perihelion distances beyond Jupiter do not significantly contribute to the inner solar system budget due to Jupiter's gravitational barrier \citep{Poppe_2016,Poppe_2019}.

Due to the sporadic nature of ZC particles in the inner solar system, a detection of any significant enhancement would require a strong concentration in semimajor axis $a$, as shown for the circumsolar dust ring co-orbiting with Venus \citep{Pokorny_Kuchner_2019}. As shown in \citet{Sommer_etal_2020}, meteoroids migrating via PR drag are likely to get captured in external MMRs of Venus and Earth and are unaffected by or quickly migrate their internal MMRs. This effect also diminishes Venus' external MMR capture rates due to the interference of internal MMRs of Earth. Nevertheless, \citet{Sommer_etal_2020} showed no or negligible efficiency of capture of any MMRs near the orbit of Mercury.

\begin{figure}
\epsscale{1.0}
\plotone{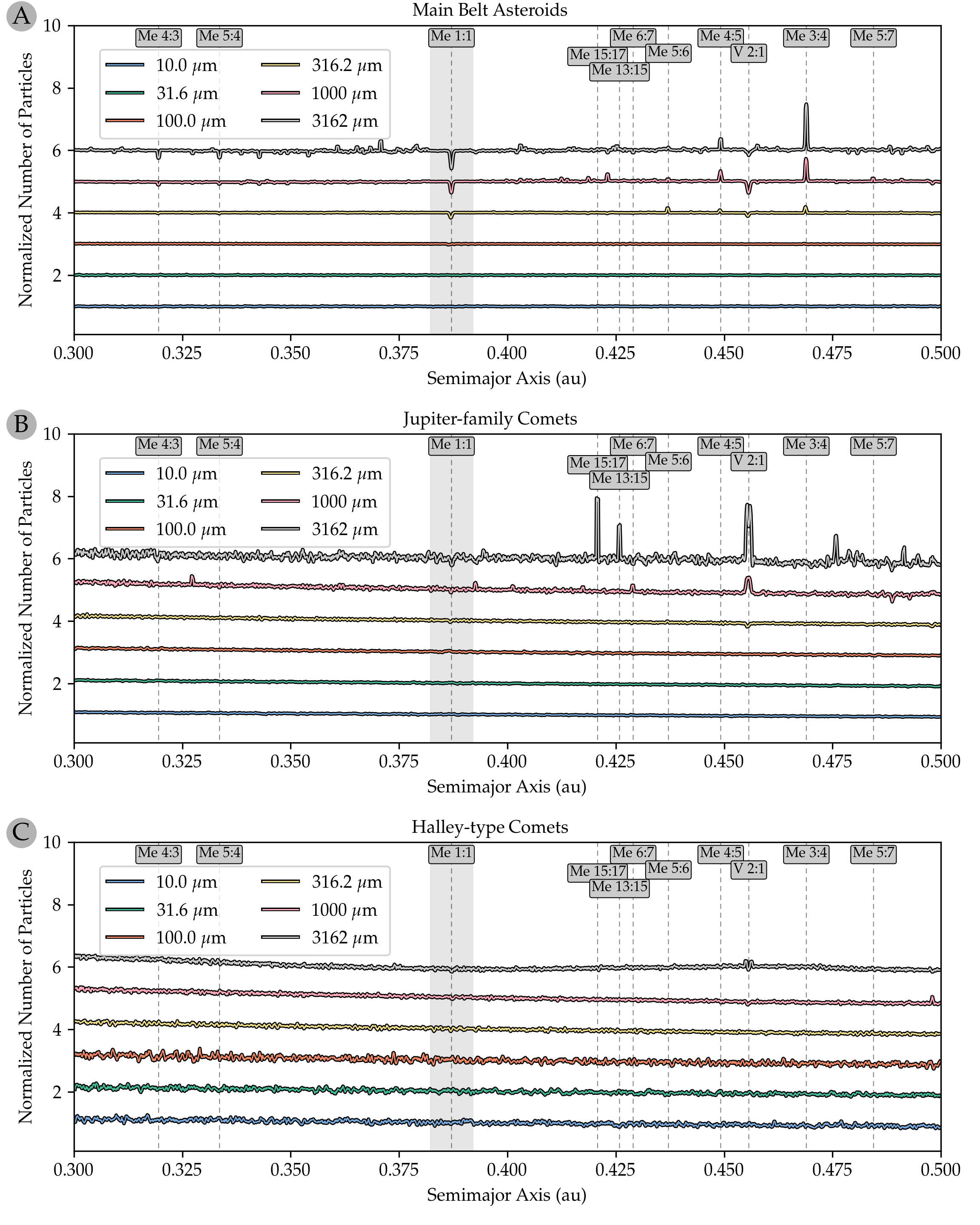}
\caption{Histograms of semimajor axes, $a$, of model particles for six different sizes and three major source populations: Panel A: Main belt asteroid meteoroids; Panel B: Jupiter-family comets; Panel C: Long-period comets. For each population, we show particles with six different diameters $D=10,~31.6,~100,~316.2,~1000,~3162~\mu$m (color coded) assuming a bulk density of $\rho=2000$ kg m$^{-3}$. The number of particles shown in the histogram is first divided by the particle semimajor axis $a$ to cancel the effect of resonance-free PR drag on the histogram slope. Then we normalize each histogram by dividing it by the median value of the histogram in the range $0.3<a<0.5$ au. We also show major mean-motion resonances with dashed vertical lines and labels at the top of each panel. There are no increased concentrations close to Mercury's orbit and only negligible clumping for $D=3162~\mu$m in several MMRs outside Mercury's orbit.}
\label{FIG:SPORADIC_HISTOGRAM_A}
\end{figure}

To test the hypothesis that the Mercury dust ring is a natural consequence of ZC particles trapped in Mercury's mean motion resonances, we expanded the numerical models used in \citet{Pokorny_Kuchner_2019, Pokorny_etal_2018, Pokorny_etal_2021}. We analyzed dust and meteoroid semimajor axis distributions near Mercury's orbit for 26 different dust/meteoroid sizes with diameters $D=0.6813~\mu$m to $D=6813~\mu$m originating from MBAs, JFCs, and LPCs, thus expanding the models used by \citet{Pokorny_Kuchner_2019, Pokorny_etal_2018, Pokorny_etal_2021}. For $D>1000~\mu$m we integrated 5,000 individual particles and for $D\le1000~\mu$m we integrated $5\times10^6/D~(\mu\mathrm{m})$ individual particles to reflect the faster particle migration via PR drag. The summary of this analysis is shown in Figure \ref{FIG:SPORADIC_HISTOGRAM_A}. For all analyzed sizes and populations, there is a maximum $\sim200\%$ increase in the number of particles for main belt and JFC meteoroids with $D\ge 1000~\mu$m related to their temporary capture in several exterior MMRs with Mercury (e.g., the 3:4 MMR for main belt asteroids and the 15:17 MMR for JFCs) and around interior 2:1 MMR with Venus for JFCs. The number of particles temporarily caught in MMRs is negligible in comparison to the total number of particles in Mercury's neighborhood and is only emphasized by the bin size of $da = 0.0003$ au. Moreover, all these temporary captures occur beyond the maximum heliocentric distance that STEREO A HI-1 can observe $R_\mathrm{hel} < 0.407$ au. For these reasons, we conclude that the sporadic meteoroid background is not a viable source of Mercury's dust ring.

\subsection{Recent asteroidal and cometary activity}
\label{SEC:RECENT_ACTIVITY}
In the previous Section, we showed that the sporadic meteoroid background does not show any signature of clumping close to Mercury's orbit. The sporadic background represents asteroidal and cometary particles that dynamically evolved for thousands to millions of years and lost their connection to their parent bodies. In this Section, we focus on freshly ejected dust and meteoroids from active asteroids and comets that can form concentrated streams of matter, meteoroid streams \citep{Jenniskens_1998}.

To analyze the potential of solar system small bodies to reproduce Mercury's dust ring, we downloaded the latest version of the MPCORB database\footnote{\url{https://minorplanetcenter.net/data} - Downloaded on April 14, 2022} which contains more than 1.2 million entries for asteroids and 942 entries for comets. For each of these bodies, we ran a minimum orbit intersection distance (MOID) analysis with respect to 360 points along Mercury's orbit using J2000 coordinates. These points were spaced uniformly in time, sampling one complete orbit of Mercury. In our analysis, we calculated the number of points on Mercury's orbit that have MOID$<0.05$ au with each object in our database. We assumed this threshold from the estimated width of Mercury's circumsolar dust ring based on the \citet{Stenborg_etal_2018} observation, where the ring was observed between solar elongations $20^\circ < \epsilon < 23^\circ$; i.e., $0.356 < R_\mathrm{hel} < 0.407$ au. We did an additional analysis with a threshold of MOID $<0.10$ au to explore a potentially more diffuse dust ring that may be beyond the field-of-view of the STEREO A HI-1 instrument. The results of our analysis showing the 5 closest asteroids, 5 closest asteroids with the absolute magnitude $H<16$ (approximately 2.2 km in diameter), and 5 closest comets are found in Table \ref{TABLE:MOID}.
\begin{deluxetable}{cccccccc}




\tablecaption{List of asteroids and comets with the highest percentage of similarity with Mercury's orbit. The table shows the five best candidates picked from all asteroids available in the MPC database, asteroids with $H<16$, and all comets in the MPC database. For each candidate, we report its name, semimajor axis ($a$) in au, eccentricity ($e$), inclination ($i$) in degrees, longitude of the ascending node ($\Omega$) in degrees, argument of pericenter ($\omega$) in degrees, percentage of Mercury's orbit within MOID of 0.05 au, and percentage of Mercury's orbit within MOID of 0.1 au. From the orbital similarity, it is evident that at least 5-10 objects with dense debris trails would be required to form the observed circumsolar dust ring close to Mercury. \label{TABLE:MOID}}

\tablenum{1}

\tablehead{\colhead{Name} & \colhead{a} & \colhead{e} & \colhead{i} & \colhead{$\Omega$} & \colhead{$\omega$} & \colhead{MOID} & \colhead{MOID} \\ 
\colhead{} & \colhead{(au)} & \colhead{} & \colhead{(deg)} & \colhead{(deg)} & \colhead{(deg)} & \colhead{$<$0.05 au(\%)} & \colhead{$<$0.10 au(\%)} } 

\startdata
\multicolumn{8}{c}{All Asteroids}\\\hline
2021 VQ3 ($H=26.2$) & 0.748 & 0.458 & 4.288 & 46.122 & 167.195 & 27.5 & 41.7 \\
2021 XA1 ($H=27.0$) & 0.723 & 0.410 & 6.294 & 65.243 & 191.885 & 27.5 & 40.8 \\
2019 XO1 ($H=24.4$) & 0.708 & 0.400 & 6.621 & 71.991 & 179.970 & 26.9 & 41.4 \\
2020 YR1 ($H=22.9$) & 0.732 & 0.420 & 8.291 & 54.827 & 235.930 & 26.7 & 39.7 \\
2017 WR ($H=24.8$) & 0.846 & 0.526 & 5.736 & 62.174 & 138.337 & 26.4 & 39.4 \\
\multicolumn{8}{c}{Asteroids with $H<16$}\\\hline
(66146) 1998 TU3   & 0.787 & 0.484 &  5.409 & 102.002 &  84.839 & 19.7 & 34.7 \\
(2212) Hephaistos  & 2.160 & 0.837 & 11.543 &  27.513 & 209.413 & 11.4 & 32.2 \\
(331471) 1984 QY1  & 2.500 & 0.893 & 14.281 & 142.268 & 337.186 & 7.8 & 30.3 \\
(164201) 2004 EC   & 1.997 & 0.860 & 34.597 &  28.824 &  10.313 & 4.4 & 14.7 \\
(68348) 2001 LO7  & 2.156 & 0.841 & 25.339 & 236.220 & 181.641 & 4.2 & 10.8 \\
\multicolumn{8}{c}{Comets}\\\hline
342P/SOHO 			& 3.042   	& 0.983 &  11.666 &  27.702 &  73.270 & 7.5 & 18.3 \\
C/2020 S3 (Erasmus) & 172.348 	& 0.998 &  19.954 & 222.735 & 350.019 & 6.7 & 20.6 \\
2P/Encke			& 2.219 	& 0.849 &  11.604 & 334.409 & 186.889 & 5.8 & 31.1 \\
323P/SOHO 			& 2.583		& 0.985 &   5.334 & 324.385 & 353.051 & 3.9 & 14.2 \\
C/2020 F3 (NEOWISE) & 341.606	& 0.999 & 129.000 &  61.031 &  37.308 & 3.6 & 7.5 \\
\enddata




\end{deluxetable}

\begin{figure}
\epsscale{1.2}
\plotone{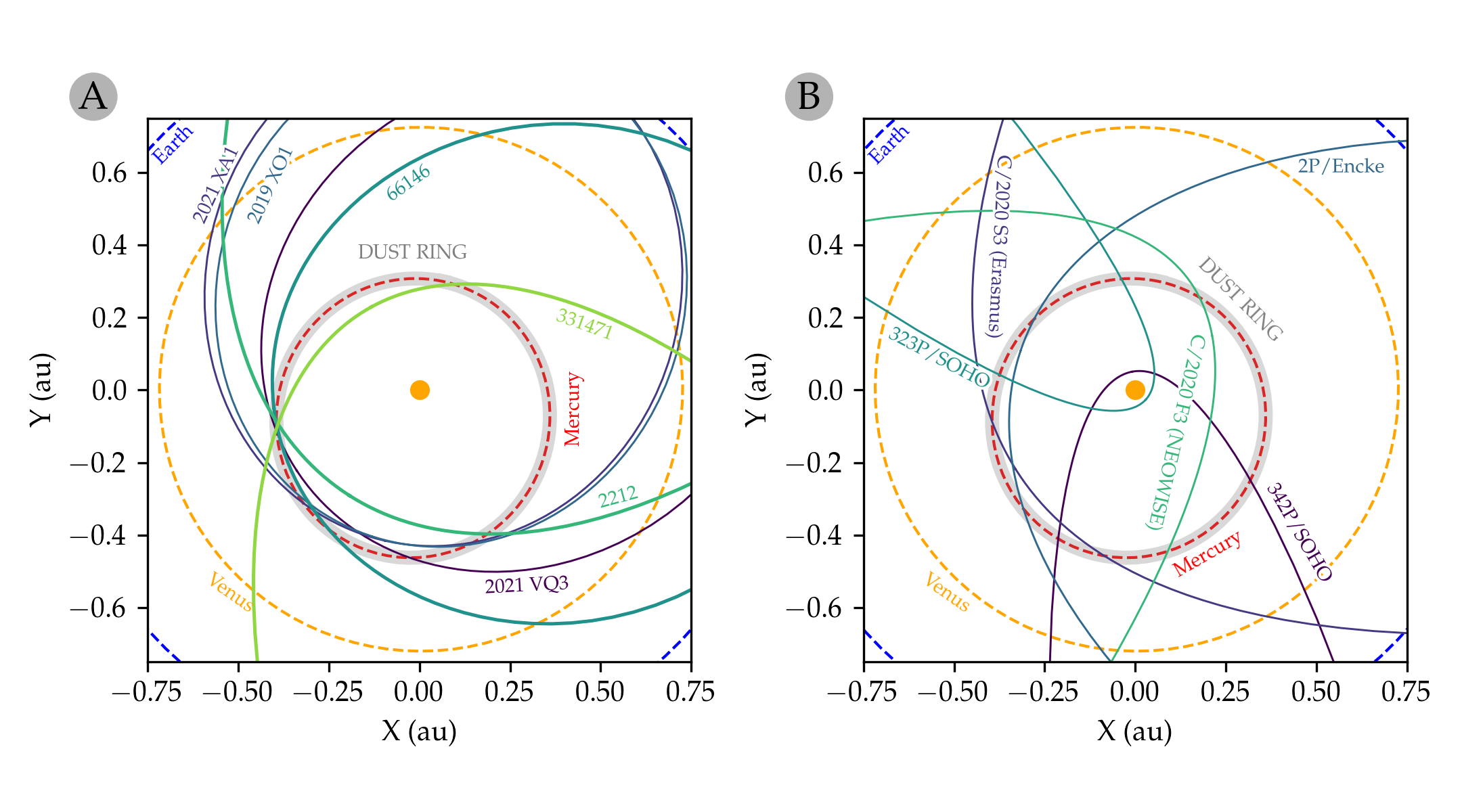}
\caption{Orbits of a selection of small bodies with the largest coverage of Mercury's orbit. Panel A: Three best candidates from all asteroids in the MPCORB database (thin solid lines) and three best candidates with $H<16$ (thick solid lines; see Table \ref{TABLE:MOID}). Panel B: Five best cometary candidates analyzed here. Both panels show the face-on view of the inner solar system; i.e., the projection on the ecliptic plane. Each small body orbit is represented by a color coded solid line and label. Orbits of Mercury (red), Venus (orange), and Earth (blue) are represented by dashed lines. Mercury's dust ring is shown as a thick gray ellipse around Mercury's orbit.}
\label{FIG:MPCORB}
\end{figure}

Results in Table \ref{TABLE:MOID} show that the smallest asteroids ($H>23$ or $D<100$~m) have the greatest similarity with Mercury's orbit. Their MOID $(<0.05 $~au) is  $\sim27\%$ so each of these asteroids' orbit overlaps with approximately one quarter of Mercury's orbit. Therefore, even if we assume that the orbits of these asteroids do not overlap close to Mercury's orbit, the current longitudinal extent of Mercury’s dust ring requires at least 3-4 of these asteroids. However, their size is orders of magnitude smaller than the amount of dust in the currently observed dust ring (Sec. \ref{SEC:Ring}). When we consider larger asteroids with absolute magnitudes $H<16$ or $D\gtrsim 2$ km (second section in Table \ref{TABLE:MOID}), the orbital coverage of Mercury's ring by individual asteroids drops significantly, requiring $>10$ objects to fully cover the longitudinal profile of the ring. This is assuming no overlap between the dust streams generated by these objects. Figure \ref{FIG:MPCORB}A, which presents a face-on view of the inner solar system and the orbits of three best candidates from all asteroids (thin solid lines) and asteroids with $H>16$ (thick solid lines), shows that this is not the case.

Similar to recent asteroid break-ups, recent comet activity or disruptions are unlikely to be the source of Mercury's dust ring. Our analysis shows that to fully cover the longitudinal structure of Mercury's dust ring, $>10$ comets would have needed to have fortuitously aligned orbits and have significant activity/established meteoroid stream (third section in Table \ref{TABLE:MOID}). As shown in Figure \ref{FIG:MPCORB}B, this is not the case. Moreover, the cometary orbits shown in Fig. \ref{FIG:MPCORB}B extend closer to the Sun with an exception of 2P/Encke, which is not observed in the STEREO observations. We also emphasize that in Fig. \ref{FIG:MPCORB} we show orbits projected into the ecliptic plane disregarding the inclination of each comet. Since most comets have inclinations significantly different from Mercury's inclination, including any latitudinal alignment condition would make the cometary origin even less probable.

Our analysis of the orbital similarities between orbits of various small bodies and Mercury's dust ring demonstrates that recent break-ups or cometary activity of known small bodies in the solar system are unlikely to be the source of Mercury's circumsolar dust ring. Even if we found several small bodies that would have a significant orbital similarity with Mercury's orbit (such as asteroids locked in the 8:7 MMR with Mercury), the dust would either need to be dynamically fresh or it would require some mechanism to keep it close to its source region. This is due to the strong effect of the PR drag that is able to change the semimajor axis of $D=1$~mm particles from 0.45 au to 0.3 au within 80,000 years. The only mechanisms that can delay the orbital decay via PR drag are the MMRs, which we investigate in the next Section.

\subsection{Hypothetical dust sources in and around the MMRs outside Mercury}
\label{SEC:MMRs}
In this Section, we analyze the ring-generating potential of dust particles released from the external MMRs of Mercury. We simulated a sample of 26 different meteoroid/dust grain diameters ranging from $D=0.6813~\mu$m to $D=6813~\mu$m with an equal logarithmic spacing. We released the particles from the vicinity of the following external MMRs of Mercury: 1:2, 2:3, 3:4, 4:5, 5:6, 6:7, and 7:8; i.e., from semimajor axis range $0.423<a<0.614$ au, where for each MMR we used its corresponding value of $a$. All particles were released with initial eccentricities of $0<e<0.2$, inclinations $0<i<10^\circ$, and randomly generated longitudes of the ascending node $\Omega$, argument of pericenter $\omega$, and mean anomaly $M$. These initial orbital elements represent values before the effect of radiation pressure is accounted for, since we want to simulate an ejection of dust grains from larger parent bodies. For each particle diameter/MMR combination, we simulated the dynamical evolution of 1,000 particles using methods described in Section \ref{SEC:Methods}. In total, we simulated the evolution of 182,000 unique particles.

\begin{figure}
\epsscale{0.97}
\plotone{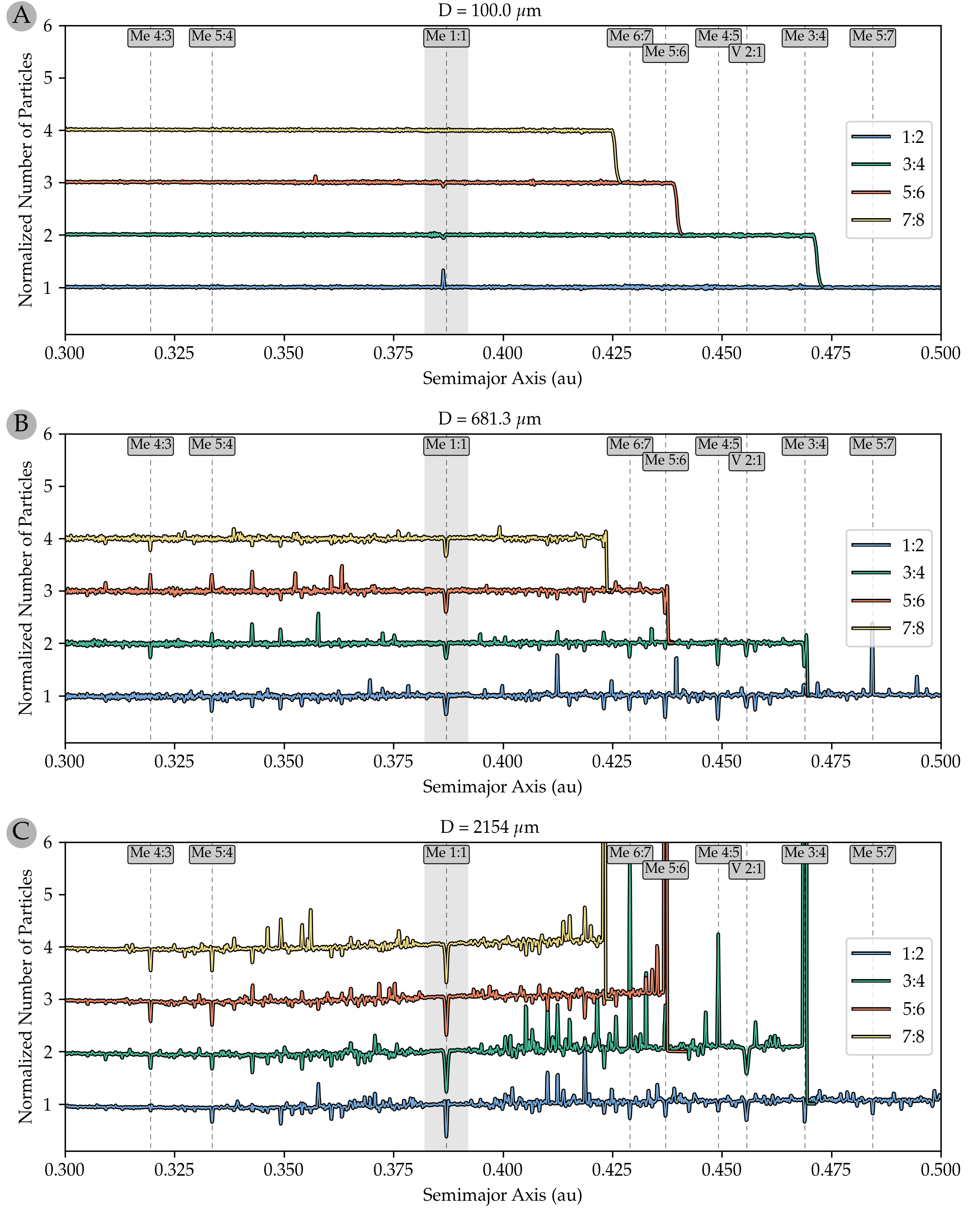}
\caption{Histograms of semimajor axes $a$ of model particles for three different diameters originating in four external MMRs of Mercury: Panel A: Particles with diameters $D=100~\mu$m; Panel B: Particles with diameters $D=681.3~\mu$m; Panel C: Particles with diameters $D=2154~\mu$m. For each particle size, we show their dynamical evolution from source regions located in four external MMRs (1:2, 3:4, 5:6, 7:8) (color coded). We also show all major mean-motion resonances with dashed vertical lines and labels at the top of each panel. We see no increased concentrations close to Mercury's orbit and only negligible clumping for $D=2154~\mu$m in several MMRs outside Mercury's orbit. Panel C also shows that larger particles $D\ge1000~\mu$m exhibit temporary stability in their source MMRs as expected due to the diminishing strength of PR drag.}
\label{FIG:Mean_Motion}
\end{figure}

The summary of contributions of source populations near these external MMRs with Mercury is presented in Fig. \ref{FIG:Mean_Motion}. Particles smaller than $D<100~\mu$m do not interact with any of the MMRs of Mercury. This is expected due to the strong effect of PR drag (Fig. \ref{FIG:Mean_Motion}A) and the large value of the semimajor axis drift $da/dt$ that allows such particles to easily skip any MMR with Mercury. As particle size increases, the semimajor-axis drift $da/dt$ due to PR drag decreases and the simulated particles spend more time ($\sim1-10$ kyr) in various resonances as shown in Fig. \ref{FIG:Mean_Motion}B for $D=681.3~\mu$m. However, this resonant capture is not efficient enough and we see no strong clumping in any of the MMRs. Moreover, particles originating in all analyzed MMRs effectively skip the 1:1 MMR with Mercury (dips at $a=0.3871$ au). For even larger particles where $D=2154~\mu$m, the particles show temporary residence in their initial MMRs (Fig. \ref{FIG:Mean_Motion}C). For the largest particle diameters analyzed here ($D=6813~\mu$m), this residence time can reach $\sim1$ million years. 

However, from the STEREO data and the stability of the longitudinal structure of the ring (Sec. \ref{SEC:Observations}), we can assume that the particle clumping necessary to reproduce the ring shape requires particles being efficiently trapped in some MMR and secular apsidal resonance with Mercury \citep{Murray_Dermott_1999_Book}. Such a capture does not efficiently occur for any combination of the MMR order and particle size that we modeled once they leave their source regions. Additionally, with increasing particle size, the particles are more likely to skip over the 1:1 MMR with Mercury, thereby creating a gap or deficiency, in contrast to the significant clustering required to explain the dust ring.

Regardless of particle size, dust and meteoroids migrating from source populations located near the external MMRs with Mercury do not get captured at or close to the 1:1 MMR with Mercury and their capture in the exterior MMRs of Mercury is inefficient. Smaller particles ($D\le100~\mu$m) smoothly drift past Mercury whereas larger particles skip the co-orbital resonance. For these reasons, we conclude that any particle population migrating from a region exterior to the 1:1 MMR with Mercury is unlikely to reproduce a circumsolar dust ring close to Mercury's orbit.

\section{Dust population generated from small bodies in and around the 1:1 MMR with Mercury}
\label{SEC:Dust_From_1:1}
In previous Sections, we showed three scenarios that do not result in particle concentrations that would be able to reproduce the structure and location of Mercury's circumsolar dust ring. From the dynamical model of Venus' circumsolar dust ring \citep{Pokorny_Kuchner_2019}, we can expect that dust particles and meteoroids released from source bodies located in 1:1 MMR with Mercury will stay temporarily locked in the MMR and will exhibit a significant amount of clustering around Mercury's orbit. There are several major dynamical differences between particles locked in co-orbital resonances with Mercury and Venus: 1) Mercury is approximately 14.7 times less massive than Venus; 2) Mercury's semimajor axis is 1.87 times shorter than that of Venus; and 3) Mercury's orbit is significantly more eccentric ($e=0.2056$) compared to Venus' almost circular orbit ($e=0.007$). Due to the smaller mass of Mercury, the strength of the 1:1 MMR is smaller \citep[e.g.,][]{Nesvorny_etal_2002}; due to the proximity of Mercury to the Sun, the semimajor-axis drift of particles in the 1:1 MMR is 1.87 times stronger than for particles co-orbiting with Venus; and due to the non-circular orbit of Mercury, we can expect more complex particle dynamics \citep{Nesvorny_etal_2002,Leleu_etal_2018}.

Similarly to Sec. \ref{SEC:MMRs}, we released dust particles and meteoroids in 32 logarithmically-spaced bins with $D=0.6813~\mu$m to $D=100,000~\mu$m to capture a full coverage of particle diameters that could produce Mercury's circumsolar dust ring. In this case, the particles were ejected from parent bodies near the 1:1 MMR with Mercury. To cover the parameter space, we released the particles from parent bodies with orbital elements selected using the following criteria: $a=0.3871$ au, $0<e<0.4$, and $0^\circ<i<14^\circ$, with randomized angles $\Omega,\omega, \mathrm{and}\, M$. We followed the particles for 100 Myr, which is an order of magnitude longer than the PR drag timescale ($T_{PR}$) for the largest particle in the sample; particles with $D=100,000~\mu$m have $T_\mathrm{PR} = 10.5$ Myr, assuming a bulk density of 2,000~kg\,m$^{3}$. The integration time step was 1 day and we recorded the particle state vectors every 100 years.

Once all particle runs were finished, we added all particles together by applying a single power-law size-frequency distribution. Since our particle sizes were log-uniformly distributed, the number of particles initially created in each size bin can be easily estimated as:
\begin{equation}
    N(D_i) = N_0 D_i^{-\alpha+1},
    \label{EQ:Number_or_Particles}
\end{equation}
where $N_0$ is a calibration constant setting the total amount of particles in the model, the index $i$ corresponds to the $i$-th bin in the size range, and $\alpha$ is the differential size index. To obtain the actual number of particles in the dynamically evolved model, we need to first divide the number of recorded particles $N_\mathrm{rec}$ by the number of initially ejected particles $N_\mathrm{init}$ and then multiply the resulting number by $N(D_i)$. This normalization is important because each model can have a different initial number of modelled particles in the simulation. 

Since remote sensing (in our case STEREO-HI1) is not sensitive to the number of particles but rather their brightness (Eq. \ref{EQ:Brightness}) or their cross-section area, $A$, we can multiply Eq. \ref{EQ:Number_or_Particles} by the particle area and get the cross-section for each size bin:
\begin{equation}
    A(D_i) = N(D_i) D_i^{2} = N_0 D_i^{-\alpha + 3}.
\end{equation}

\begin{figure}
\epsscale{1.2}
\plotone{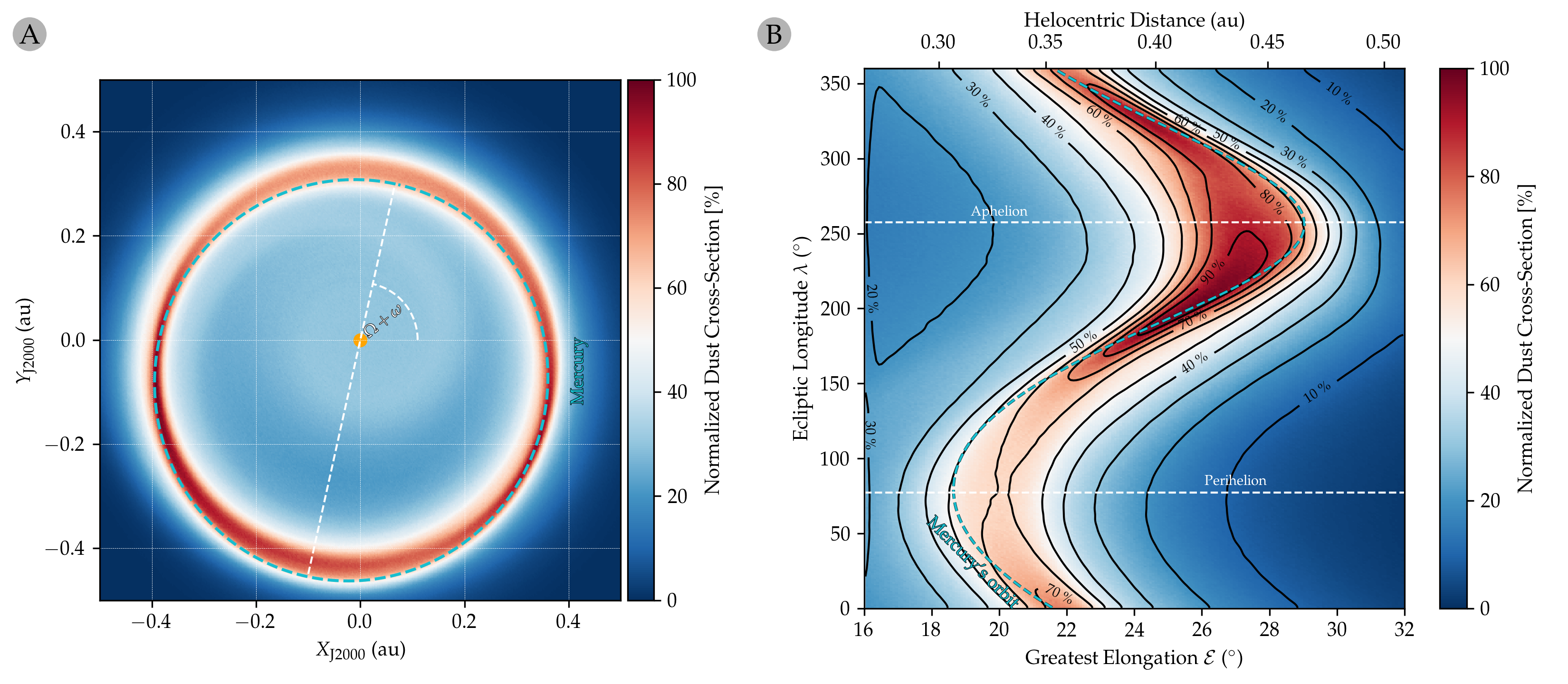}
\caption{Panel A: The face-on view of dynamically evolved dust released from parent bodies locked in the 1:1 MMR with Mercury. The color shows the normalized dust cross-section $A$. Mercury's orbit is shown as a dashed cyan ellipse. The line of apses is represented by the white dashed line. Panel B: Longitudinal variations of our model for Mercury's dust ring. The primary $x-$axis shows the greatest elongation $\mathcal{E} = \mathrm{arcsin}(r_\mathrm{hel}/0.96)$, the secondary $x-$axis shows  heliocentric distance in au, and the $y-$axis the ecliptic longitude measured from the vernal point. The color-coding shows the normalized dust cross-section of our model where the contours show 10\% increments. The cyan dashed line represents Mercury's orbit. }
\label{FIG:MMR_Cross_Section}
\end{figure}

\begin{figure}
\epsscale{0.6}
\plotone{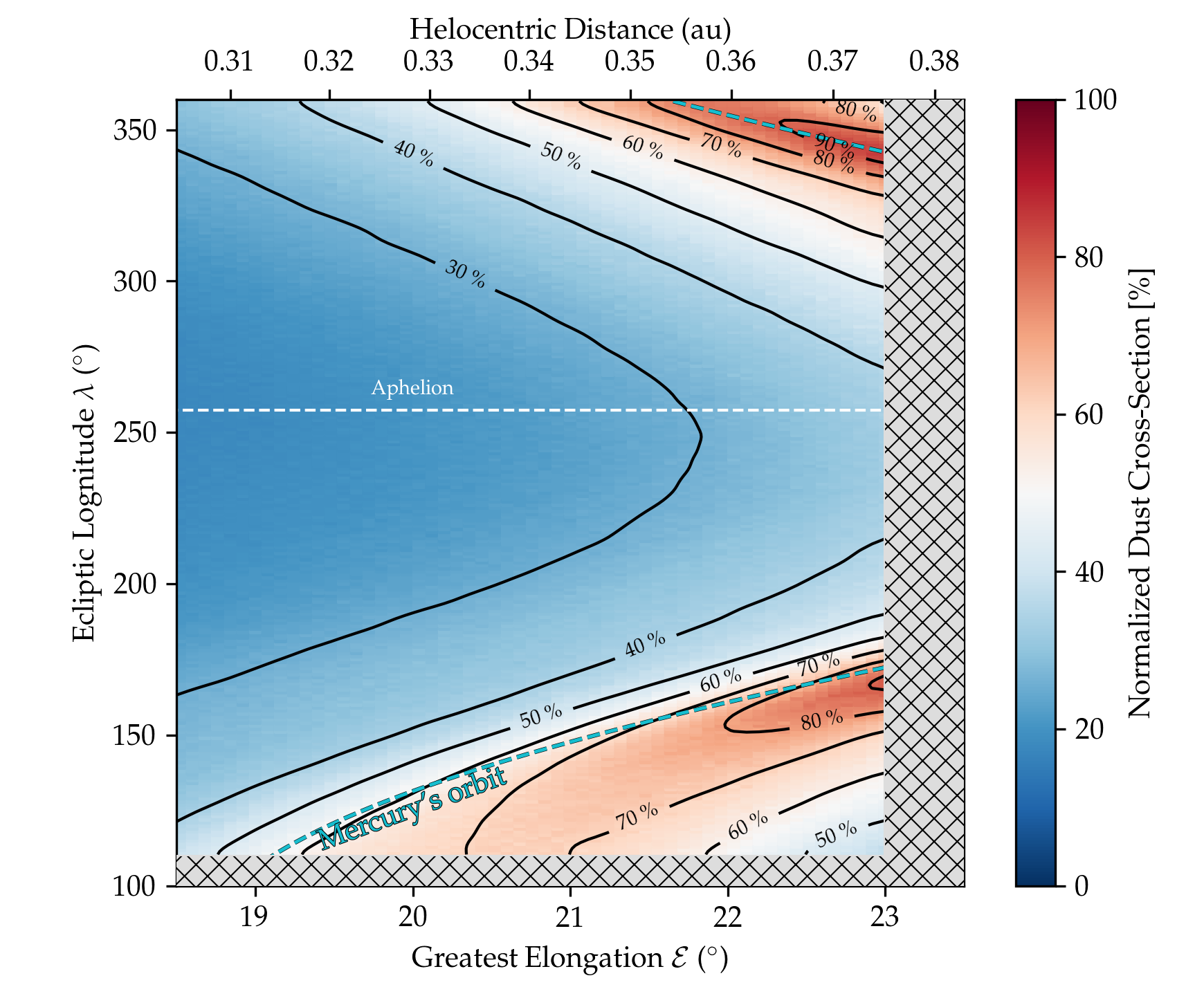}
\caption{The same as Fig \ref{FIG:MMR_Cross_Section}B, but now zoomed in to the extent of the \citet{Stenborg_etal_2018} observation shown in Fig. \ref{FIG:STENBORG_DATA}. The main difference between our model and the observed dust ring is the lack of dust around $\mathcal{E}=22.5^\circ$ and $\lambda= 115^\circ$. We attribute this difference to an additional source of dust originating from 2P/Encke (Fig. \ref{FIG:Encke_Contribution}). }
\label{FIG:MMR_Cross_Section_Zoom}
\end{figure}

Figure \ref{FIG:MMR_Cross_Section}A shows the face-on view of the distribution of dust cross-section $A$ of particles released from bodies near the 1:1 MMR. Our results clearly show that our model produces an eccentric ring of dust that is aligned with the orbit of Mercury. To combine millions of years of simulated dynamical evolution at each recorded time step, we rotate the reference frame of the recorded particle state vector by angle $-\varpi+\varpi_\mathrm{J2000}$ along the $z$-axis, where $\varpi$ is Mercury's instantaneous longitude of the periapsis and $\varpi_\mathrm{J2000}$ is the initial (J2000) value for Mercury. Since the dust ring in Fig. \ref{FIG:MMR_Cross_Section} evolved for millions of years and the non-relativistic precession period of Mercury's perihelion is $\approx243,500$~yr, the shape of the dust ring is stable on million-year timescales and its time evolution is mostly driven by radiation forces from the Sun. To better show the longitudinal variations of the model dust ring, we calculated the ecliptic longitude of each particle $\lambda = \mathrm{atan2}(y,x)$ and its greatest elongation $\mathcal{E} = \mathrm{arcsin}(r_\mathrm{hel}/0.96)$ (Fig. \ref{FIG:MMR_Cross_Section}B). Here, atan2 is the 2-argument arctangent that removes ambiguity in the Cartesian-to-polar coordinate conversion, and the value of 0.96~au is the heliocentric distance of STEREO A. Fig. \ref{FIG:MMR_Cross_Section}B shows that most of the dust cross-section is concentrated close to Mercury's aphelion, which is expected due to the smaller orbital velocities of dust particles in this region. For easier comparison with the data from STEREO A observations, we extracted the same region and show it in Figure \ref{FIG:MMR_Cross_Section_Zoom}. Though not directly comparable due to brightness scaling with heliocentric distance, we see a good match of the modeled ring longitudinal profile and the dust cross-section variations to the dust brightness excess seen in Fig. \ref{FIG:STENBORG_DATA}. Both plots show results using the SFD index $\alpha=3.5$. We tested various values of alpha $\alpha\in[2.0,5.0]$ and the dust ring structure remains unchanged.

While the first results are optimistic, there are several additional steps that we need to assess before we can call our hypothesis plausible. In the following Sections, we use our simple STEREO simulator to model the relative brightness increase in our model and compare it to the STEREO data. Then we search for a potential source of this dust ring, as currently no obvious sources are readily available.

\subsection{Comparison to STEREO data}
\label{SEC:Model_Comparison_To_Stereo}
In the previous Section, we showed that dust and meteoroids released from source bodies locked in the 1:1 MMR with Mercury produce an eccentric circumsolar ring of dust that is aligned with Mercury's orbit, and its shape is stable over millions of years. To compare our results to available data, we use the simple STEREO A simulator described in Sec. \ref{SEC:Simulator}. As shown in Fig. 4 and Sec 3.1.2 in \citet{Stenborg_etal_2018}, STEREO A observes light scattered by the dust ring particles at longitudes shifted by $\approx70^\circ$ because the line of sight tangentially traverses the ring at elongations $\epsilon \approx 20^\circ$. Therefore, STEREO A observing at ecliptic longitude $\lambda = 340^\circ$ detects the strongest brightness increase from the ring sections with ecliptic longitudes $\lambda = 270^\circ$. The HI-1 instrument is pointing in the eastward direction (behind the spacecraft with respect to its direction of motion). The westward pointing detector would experience a similar shift but in the positive direction. To accommodate this $70^\circ$ shift in $\lambda$ and to make it easier to compare to our dynamical model from Sec. \ref{SEC:Dust_From_1:1}, we use $\lambda = (\lambda_\mathrm{S/C} - 70^\circ)$ in all subsequent Figures in this article.

To obtain a relative brightness increase of our model ring with respect to the ZC and data in Fig. \ref{FIG:STENBORG_DATA}, we perform two additional operations with our brightness model. First, we multiply the brightness of each particle in our model $\mathcal{B}$ by a factor $\epsilon^{2.35}$ to capture the brightness increase of the smooth ZC background with $\epsilon$ as discussed in \citet{Stenborg_etal_2018_Fcorona}. We also tested different scaling factors from $\epsilon^{2.30}$ up to $\epsilon^{2.35}$ and our analysis yielded similar results. Second, we subtract the median brightness at elongation $\epsilon=18^\circ$ from the entire model to remove the contribution of the dust that blends in with the smooth ZC background to replicate the approach of \citet{Stenborg_etal_2018}. We set all negative values of the relative brightness increase after subtracting the median value to zero.

\begin{figure}
\epsscale{1.2}
\plotone{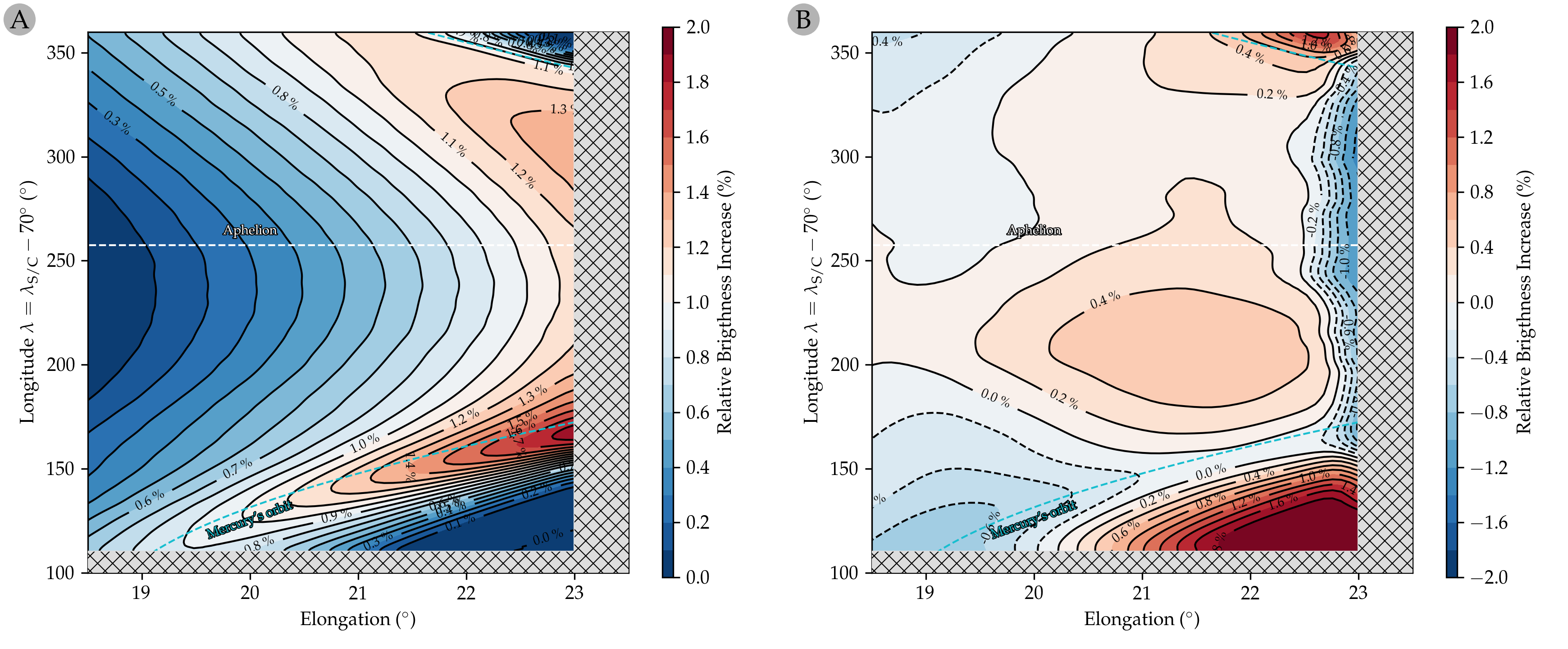}
\caption{Panel A: Relative brightness increase of the best fit model of dust originating in 1:1 MMR with Mercury. The plot shows the same region as the \citet{Stenborg_etal_2018} observation (Fig. \ref{FIG:STENBORG_DATA}) where the $x-$axis is the solar elongation $\epsilon$ and the $y-$axis is the spacecraft longitude shifted by $-70^\circ$ to account for the observing geometry (see text for explanation). Panel B: Model residuals of our best fit model. Our model does not reproduce the relative brightness increase in the bottom right corner of the Figure, that can be attributed to dust from comet 2P/Encke.}
\label{FIG:Model_STEREO_Comparison}
\end{figure}

Figure \ref{FIG:Model_STEREO_Comparison}A shows the best fit of our dust ring model to the data and \ref{FIG:Model_STEREO_Comparison}B shows the model residuals. The quality of the fit was determined by minimizing the root-mean-square deviation where the free parameters were the size-frequency distribution index $\alpha_\mathrm{ring}$ and the maximum relative brightness increase of the ring $F_M$. The best fit was achieved for $\alpha_\mathrm{ring} = 4.03$ and $F_M = 1.93$. Considering the fact that the observational data set from \citet{Stenborg_etal_2018} is longitudinally averaged with a $40^\circ$ and consists of 6+ years of data, we find that the model fits the ring well. There are two features that our current model cannot reproduce: 1) an enhancement in brightness located at $\epsilon =23^\circ$ and $\lambda = 360^\circ$ that could be attributed to the longitudinal averaging, and 2) a broad enhancement at $\epsilon = 22.5^\circ$ and $\lambda = 120^\circ$ that is attributed  to dust from comet 2P/Encke. We explore the contribution of 2P/Encke in the following Section.

\subsection{Contribution of 2P/Encke to Mercury's dust ring}
\label{SEC:2PEncke}
Figure \ref{FIG:Model_STEREO_Comparison} showed that our model is unable to reproduce a significant relative brightness increase located at $\epsilon = 22.5^\circ$ and $\lambda = 120^\circ$. \citet{Stenborg_etal_2018} suggested that 2P/Encke's orbit is in a favorable configuration that could explain this enhancement. Moreover, models of impact-driven calcium exosphere at Mercury show that 2P/Encke plays a major role in the calcium production rate when Mercury is close to its pericenter \citep{Killen_Hahn_2015,Christou_etal_2015,Pokorny_etal_2018}. To explore the potential 2P/Encke contribution, we created a simple dynamical model. First, we backward integrated 1000 clones of 2P/Encke for 5,000 years, where the orbital elements and the covariance matrix were obtained from the JPL Small-Body Database Lookup API for body: 2P/Encke [2015] (K235/2) - default 2\footnote{\url{https://ssd-api.jpl.nasa.gov/sbdb.api?sstr=2P&cov=mat&full-prec=true}}. The 5,000 year time span is based on the conclusions of \citet{Egal_etal_2021_dynamical} that the 2P/Encke meteoroid stream (Taurid complex) might be the result of the fragmentation of a larger body 5,000-6,000 years ago using orbit convergence methods. The clones were generated using a standard multivariate normal distribution available in NumPy \citep{Numpy}. Then we simulated the dynamical evolution of particles with diameters from $D=10.0~\mu$m to $D=4642~\mu$m (in 9 logarithmically spaced bins), assuming bulk density $\rho=2000$ kg m$^{3}$, released in 100-year intervals and recorded their positions in the year 2020. We used the same integration methods as in the previous Sections described in Sec. \ref{SEC:Methods}.

To obtain data comparable to the observations by \citet{Stenborg_etal_2018}, we adopted the same methods described in Sec. \ref{SEC:Model_Comparison_To_Stereo}; i.e., we normalized the brightness of each particle by its elongation and then subtracted the median value at $\epsilon=18^\circ$. Furthermore, we included an additional filter for the latitudinal extent of the STEREO field-of-view. \citet{Stenborg_etal_2018} does not provide any information about the latitudinal extent of the ring due to the fact that the nature of their measurement \textit{"precludes an analysis of its latitudinal extent"}. The STEREO HI-1 instrument has an angular FOV of $20^\circ$ \citep[Fig.  4 in][]{Eyles_etal_2009} and thus we apply a $8^\circ$ cutoff for the maximum absolute value of the ecliptic latitude of particles that contribute to our model. The $8^\circ$ corresponds to the height of a triangle with a $10^\circ$ hypotenuse and $6^\circ$ base; the STEREO FOV is centered at $\epsilon = 14^\circ$ and the ring structure is $\approx 6^\circ$ away. Contributions of particles of all sizes are combined using a single power-law size-frequency distribution (Eq. \ref{EQ:Number_or_Particles}) with the differential size index $\alpha_\mathrm{comet}$. To obtain the best model fit for just the cometary contribution, we fit only the region between $110^\circ<\lambda<135^\circ$ and $21.4^\circ < \epsilon < 23.0^\circ$.

\begin{figure}
\epsscale{1.2}
\plotone{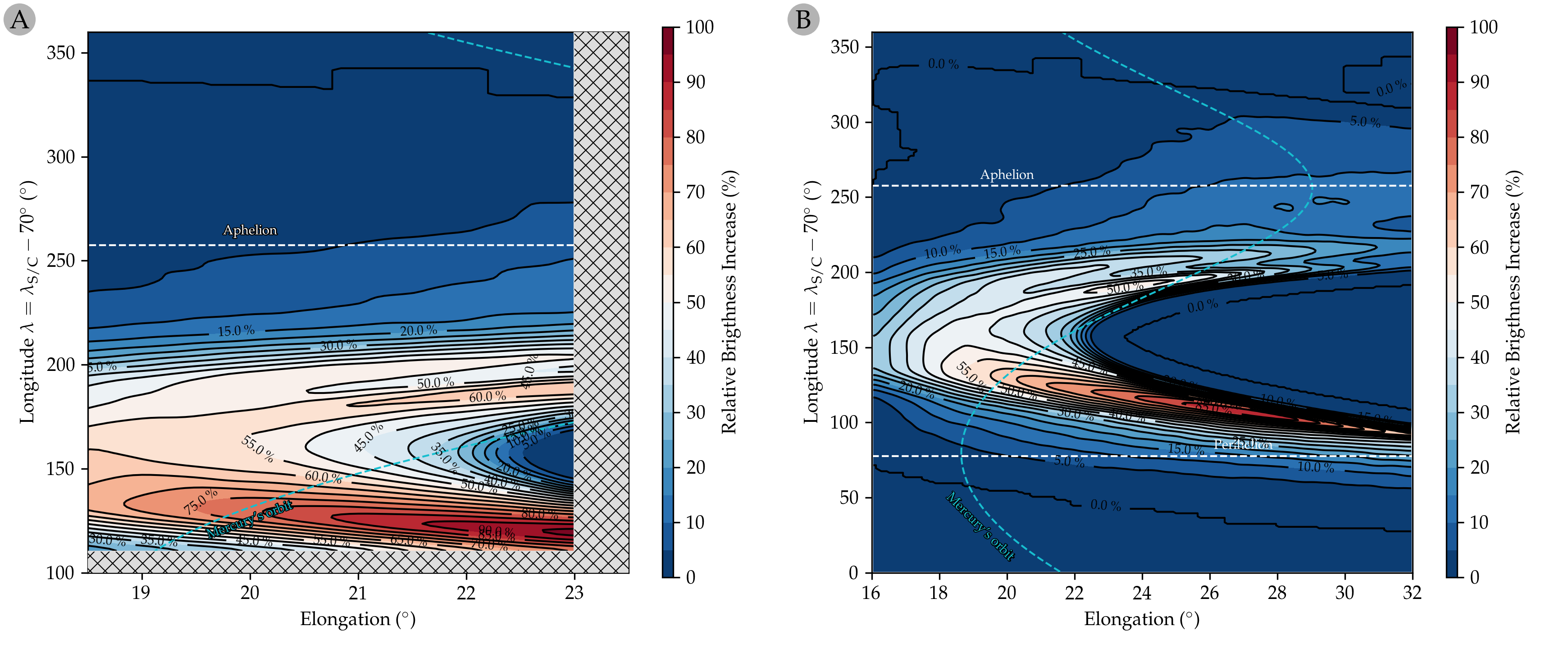}
\caption{Panel A: The same as Fig \ref{FIG:Model_STEREO_Comparison}A but now for our model of the 2P/Encke meteoroid stream. The relative brightness increase reaches its maximum at $\epsilon=23^\circ$ and $\lambda=114^\circ$, which is a favorable location for the portion of the observation that our dust ring model does not reproduce. However, the cometary contribution spans a wide range of longitudes and contributes significantly to regions where our ring model fits the data well. Panel B: A wider extent of the relative brightness increase of our 2P/Encke model. The dust stream is aligned with the longitude of periapsis of 2P/Encke $\varpi = 161^\circ.1$. Note that the longitude in this Figure is comparable to the heliocentric ecliptic longitude, where $\lambda = \lambda_\mathrm{S/C} - 70^\circ$ reflects the $70^\circ$ shift due to observational geometry.}
\label{FIG:Encke_Contribution}
\end{figure}

\begin{figure}
\epsscale{1.2}
\plotone{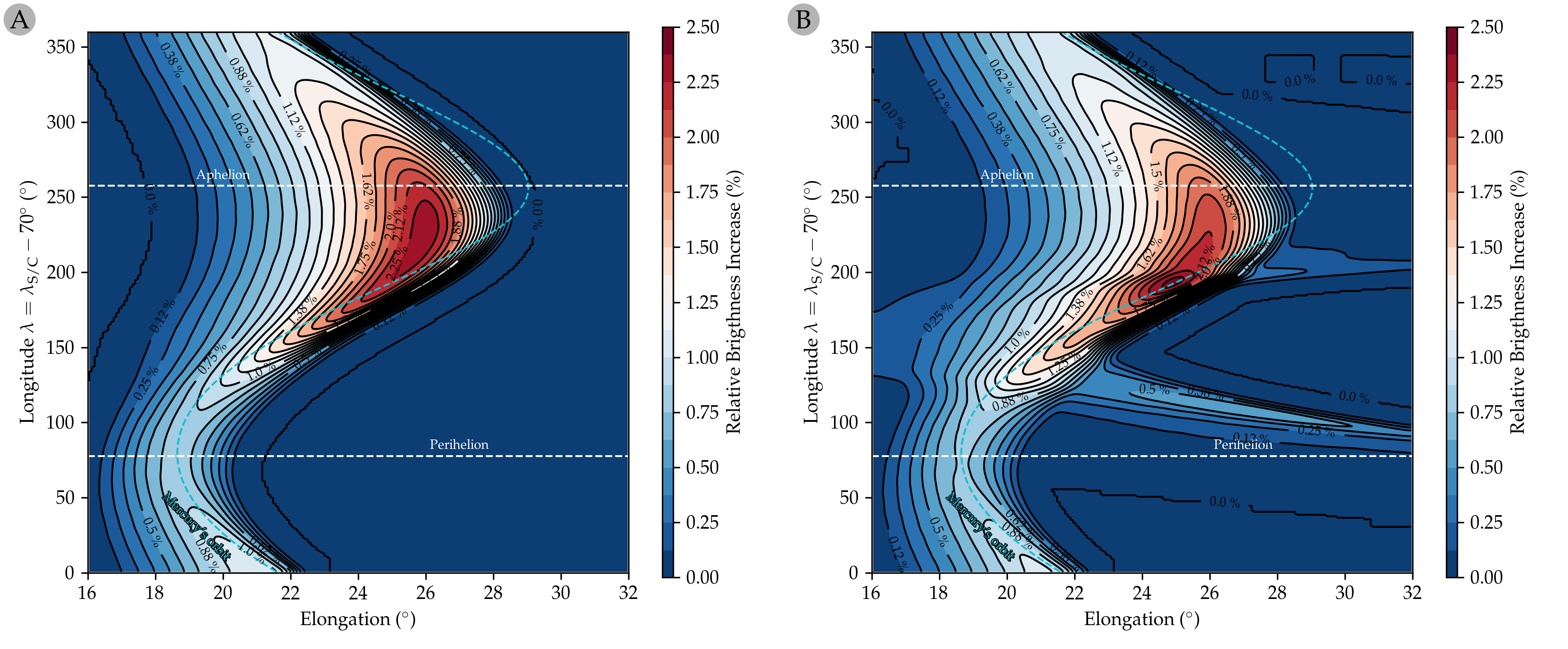}
\caption{Panel A: The same as Fig \ref{FIG:Encke_Contribution}B, but now for our model of the dust ring originating from particles co-orbiting with Mercury shown in Section \ref{SEC:Model_Comparison_To_Stereo}. The dust ring signature extends up to $\epsilon\approx28^\circ$ and its observations could provide additional constraints for Mercury's dust ring. Panel B: The same as Panel A but for the best fit model that combines our dust ring model and the contribution of 2P/Encke. This view shows that the magnitude of the relative brightness increase of 2P/Encke would be easy to constrain if the observation extended beyond $\epsilon = 24^\circ$ for $100^\circ < \lambda < 150^\circ$ where we expect no signal from the dust co-orbiting with Mercury. Note that the longitude in this Figure is comparable to the heliocentric ecliptic longitude, where $\lambda = \lambda_\mathrm{S/C} - 70^\circ$ reflects the $70^\circ$ shift due to observational geometry.}
\label{FIG:Ring_Comet_Comparison}
\end{figure}

Figure \ref{FIG:Encke_Contribution}A shows our best fit of the 2P/Encke model to the selected data region using a differential size index of $\alpha_\mathrm{comet} = 3.02$. The cometary contribution to the relative brightness increase is asymmetric in $\lambda$ with respect to the longitude of periapsis of 2P/Encke $\varpi = 161^\circ.1$ due to the latitudinal filter that we applied to simulate the STEREO HI-1 FOV and the fact that a portion of particles seen at $\lambda>150^\circ$ is outside the FOV. When neglecting the latitudinal extent filter, the asymmetry of the brightness increase is negligible. This asymmetric shape of the brightness increase is a potential solution for filling the gap that our model for Mercury's dust ring cannot reproduce (Sec. \ref{SEC:Model_Comparison_To_Stereo}). However, as shown in Fig. \ref{FIG:Encke_Contribution}A, the contribution to $\lambda>150^\circ$ is not negligible and is $\sim50\%$ of the maximum relative brightness increase in the model. Therefore, filling the gap by combining our dust ring model from the previous section and our 2P/Encke model does not result in a significantly better fit. Either the comet's contribution is too strong and distorts the rest of the ring, or the comet provides only a fraction of the missing signal at $110^\circ<\lambda<135^\circ$ and $21.4^\circ < \epsilon < 23.0^\circ$. The best model fit combining both the ring and comet models has the following parameters: $\alpha_\mathrm{ring} = 4.28, \alpha_\mathrm{comet} = 2.71, F_\mathrm{M(ring)}=1.74, F_\mathrm{M(comet)=0.49}$, where the goodness-of-the-fit increases only by 7\%. We show this best fit model in Fig. \ref{FIG:Ring_Comet_Comparison}B and provide a direct comparison with the dust ring only model in Fig. \ref{FIG:Ring_Comet_Comparison}A.

Figure \ref{FIG:Encke_Contribution}B shows the 2P/Encke relative brightness increase for a wider range of elongations $\varepsilon$ and the entire range of longitudes $\lambda$. The 2P/Encke stream extends beyond the dust sourced by bodies in the 1:1 MMR with Mercury and should be detectable by remote sensing similar to the discovery of Mercury's dust ring \citep{Stenborg_etal_2018}. Such an observation would provide an independent constraint for determining the contributions of the dust co-orbiting with Mercury and the 2P/Encke meteoroid stream.

Our model for 2P/Encke was simplified for computational reasons as we originally thought that it is a peculiar effect intended for future work. Recently, comprehensive models for the 2P/Encke meteoroid stream and its effect on Earth observed as the Taurid complex were published by \citet{Egal_etal_2021_dynamical} and \citet{Egal_etal_2022}. \citet{Egal_etal_2022} shortly discuss impacts of 2P/Encke meteoroids on Mercury, but do not discuss their potential signature in STEREO data. We will seek the implementation of these new comprehensive models in future work.

\subsection{Can the small body population be primordial?}
\label{SEC:Residence_Time}
Previous Sections showed that the only potential source of the currently observed brightness increase is a population of dust particles locked in the 1:1 MMR with Mercury. \citet{Pokorny_Kuchner_2019} showed that Venus co-orbital asteroids could be  primordial and survive 4.5 Gyr in the 1:1 MMR with Venus. However, the search for such asteroids has been unsuccessful so far \citep{Pokorny_Kuchner_Sheppard_2020, Sheppard_etal_2022} and \citet{Pokorny_Kuchner_2021} showed that the Yarkovsky force can hinder the primordial origin hypothesis. Mercury, being closer to the Sun and less massive than Venus, is not the most viable candidate for keeping its primordial co-orbital small body population. Nevertheless, we decided to test the primordial origin hypothesis as well as the dynamic stability of particles of various sizes by analyzing the co-orbital residence time of test particles of different sizes with diameters from $D=0.1$ mm to $D=100$ mm using the same simulation scheme described in Sec. \ref{SEC:Methods}; i.e., the PR drag and the solar wind are added as additional drag forces. Additionally, we performed a simulation with the gravity only to obtain the maximum residence time for any small body in the 1:1 MMR with Mercury. The co-orbital residence time $T_\mathrm{res}$ is calculated as the time when the number of particles in the 1:1 MMR decreased below 2\% of the initial number of simulated particles. To determine if the particles are in the 1:1 MMR, we evaluate whether the particle semimajor axis $a$ is within 1\% of Mercury's instantaneous $a_\mathrm{Mer}$ in the simulation (e.g,$0.3832<a<0.3910$ au for $a_\mathrm{Mer} = 0.3871$ au). Note that the half-width of the 1:1 MMR with Mercury is $0.38\%$ \citep{Murray_Dermott_1999_Book} and using this cut-off provides almost identical results to our simulation.

\begin{figure}
\plotone{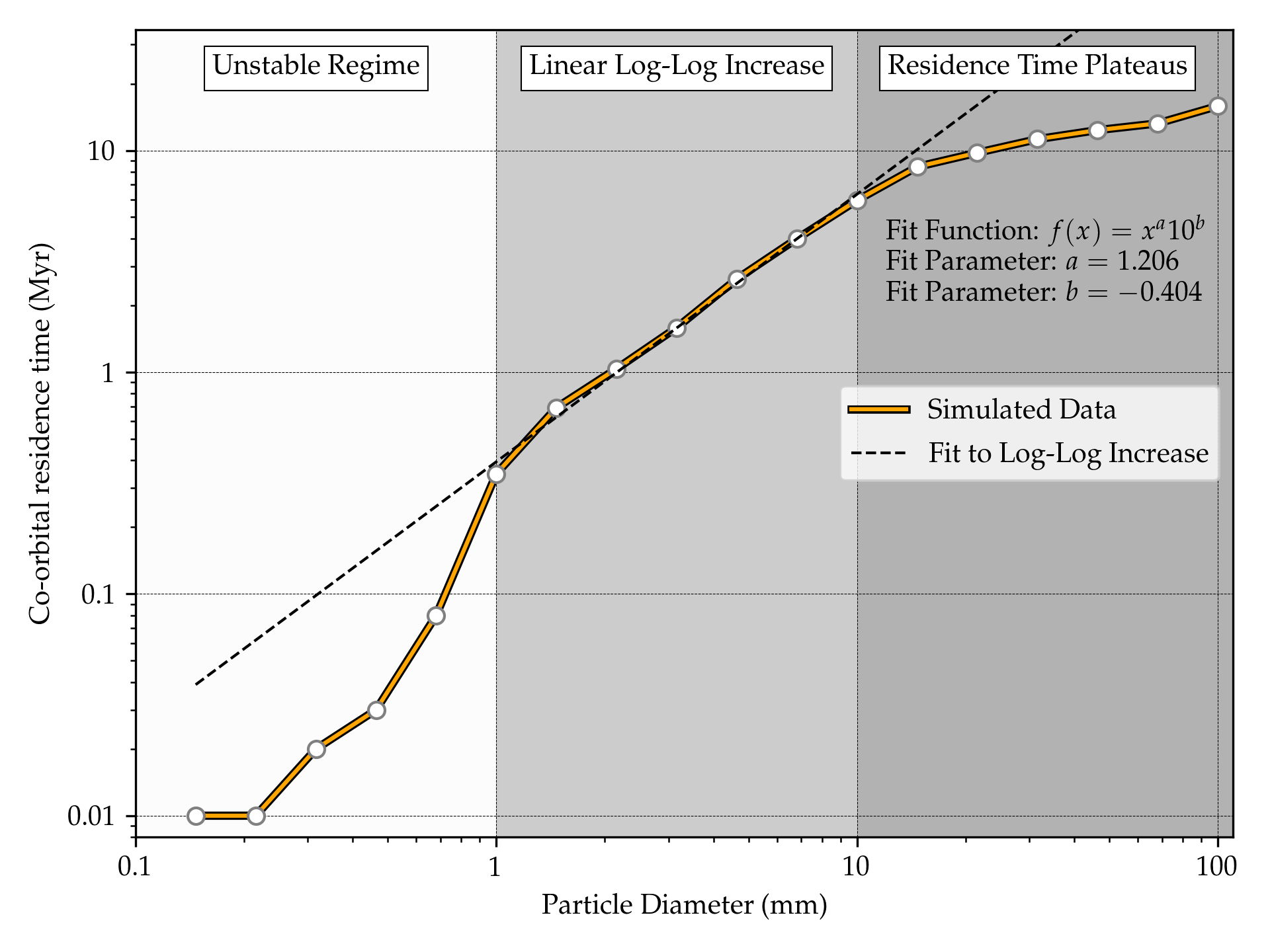}
\caption{The co-orbital residence time $T_\mathrm{res}$ with respect to simulated particle diameter $D$ (orange line with circles). Both $x$- and $y$-axes are on the logarithmic scale. The greyscale boxes delineate different regimes of $T_\mathrm{res}$ behavior. We fitted the $1\le D<10$ mm range with a power law (dashed black line), $f(x)=x^a10^b$, where $x=D$ is the particle diameter, $a$ is the power in the power law, and $b$ provides the offset.  }
\label{FIG:Residence_Time}
\end{figure}

Figure \ref{FIG:Residence_Time} shows that the co-orbital residence time $T_\mathrm{res}$ increases with particle diameter $D$ and ranges from thousands of years to tens of millions of years. We identified three different regimes of $T_\mathrm{res}$ behavior depending on $D$: a) the unstable regime for $D<1$ mm, where the drag forces are too strong for particles to get efficiently captured and particles migrate quickly from the 1:1 MMR; b) the linear log-log increase for $1 \le D < 10$ mm, showing $T_\mathrm{res}\propto D^{1.2}$ behavior that is expected from the decreasing efficacy of PR drag with increasing $D$; and c) the residence time plateau for $D\ge10$mm, where $T_\mathrm{res}$ stops following the exponential growth with increasing $D$ and plateaus at around 15 Myr. Our test without additional drag forces showed $T_\mathrm{res~(D = \infty)} = 16.3$ Myr, which is comparable to our $D=100$ mm particle simulation ($T_\mathrm{res(D=100~\mathrm{mm})} = 15.9$ Myr).

From our analysis we conclude that no primordial small body is able to survive more than 20 Myr in the 1:1 MMR with Mercury and thus the source of the current ring must be provided by a rather recent event due to the maximum co-orbital residence time $T_\mathrm{res}<20$ Myr. 
\citet{Tabachnik_Evans_2000} analyzed the stability of Mercury's co-orbitals and for some orbital configurations the co-orbital stability in horseshoe orbits extended to 100 Myr. This would open a longer windows for a recent impact but still prevents a primordial origin of the dust ring.
We calculated that the currently observed dust would be equivalent to a several-km-diameter sphere of material, which leads to two potential sources: 1) a captured asteroid, and 2) a recent impact on Mercury. \citet{Greenstreet_etal_2020} showed that the capture of Centaur asteroids into 1:1 MMR with Jupiter is possible but inefficient, where approximately 1 in 100,000 asteroids gets temporarily captured. They also find that the maximum residence time is $<100,000$ years. Since Mercury is $\sim 5700$ times less massive than Jupiter, we assume that the resonant capture efficiency is negligible and not viable as a potential progenitor of Mercury's dust ring. Therefore, to support the recent impact hypothesis, we analyze whether a recent impact on Mercury is able to transport the material into 1:1 MMR with Mercury and maintain the ring structure until today.

\subsection{Recent impact on Mercury can transfer ejecta into 1:1 MMR with Mercury}
\label{SEC:RecentImpact}
Mercury's surface shows definite markings of small body impacts in the geologically recent past. \citet{Kinczyk_etal_2020} identified 22 "Class 5" craters with diameters larger than $D>40$ km, with two of these craters being larger than 100 km in diameter. Class 5 craters are the freshest craters classified on Mercury's surface, and are interpreted to have formed during the Kuiperian cratering era, which spans from $<300$ Ma through today \citep{Banks_etal_2017}. This could potentially mean that some of the freshest Class 5 craters have ages $<20$ Ma, i.e., our estimated lifespan of Mercury's circumsolar ring (Sec \ref{SEC:Residence_Time}). To evaluate whether the formation of any of these young craters could explain the observed dust ring, we first estimate the mass of material that can be transferred from the surface into the co-orbital resonance with Mercury during an impact event.

The amount of ejecta released during cratering events and the ejecta velocity distribution are not well constrained \citep{Holsapple_1993,Holsapple_Housen_2007}. Due to the size of the craters ($D>1$ km), the impact-related calculations are strictly in the gravity regime \citep{Holsapple_1993}.
Let us assume that an impactor with a radius $r_\mathrm{imp}$, bulk density $\delta$, and  the normal impact velocity component $U$ impacts Mercury's surface.
Using the relations from \citet{Holsapple_Housen_2007} we can estimate the mass of ejecta $M_e$ produced by a point source impactor made of sand or cohesive soil with velocity $v$ where the ejected mass has ejecta velocity $v_\mathrm{ej}$ larger than $v$ as
\begin{equation}
    M_e(v_\mathrm{ej}>v) = m_\mathrm{imp} 0.018\left(\frac{v}{U}\right)^{-1.23} \left(\frac{\delta}{\rho} \right)^{0.2},
    \label{EQ:Ejecta_Mass}
\end{equation}
where $m_\mathrm{imp}$ is the impactor mass, and $\rho$ is the impacted surface density. To estimate the impactor mass, we can employ the estimated crater radius $R_c$ caused by our model impactor from \citet{Holsapple_Housen_2007}:
\begin{equation}
    R_c = r_\mathrm{imp} 1.03 \left( \frac{g r_\mathrm{imp}}{U^2} \right)^{-0.17} \left( \frac{\delta}{\rho} \right)^{0.332},
    \label{EQ:Crater_Radius}
\end{equation}
where $g=3.7$ m s$^{-2}$ is the gravitational acceleration on the surface of Mercury. Let us assume that the impactor and Mercury's surface have similar densities and $\delta/\rho \approx 1$. Then for an impactor with $U = 30,000$ m s$^{-1}$ \citep[e.g.,][]{Marchi_etal_2009}, $\delta=3,000$ kg m$^{-3}$ and the resulting crater diameter 40 km we obtain the impactor radius $r_\mathrm{imp}=2,811$ m and $m_\mathrm{imp}=2.79\times10^{14}$ km. The amount of ejecta with ejection velocities larger than Mercury's escape velocity $v_\mathrm{ej} > 4220$ m s$^{-1}$ is $M_e(v_\mathrm{ej}>4220\mathrm{~m s}^{-1})=5.60\times10^{13}$ kg, which translates into a sphere of radius of 1646 meters. Eqs. \ref{EQ:Ejecta_Mass} and \ref{EQ:Crater_Radius} show that for a fixed crater radius $R_c$, the mass ejected is not proportional to the tangential impact velocity $U$ because $m_\mathrm{imp}\propto U^{1.23}$ and the impact velocity term cancel out in Eq. \ref{EQ:Ejecta_Mass}.

Continuing the impact example in the previous paragraph, after a recent impact of a $\sim3$ km object, approximately $1.25\times10^{14}$ kg of material reaches the boundary of Mercury's Hill sphere. Ejecta released with velocities $v_\mathrm{ej}<4,220$ km s$^{-1}$ do not reach the Hill sphere boundary ($r=0.0012$ au or $r=0.1753 \times 10^6$ km) and likely fall back to the planet. We look for ejecta that reach Mercury's Hill sphere with a heliocentric velocity that puts the ejected particle into 1:1 MMR with Mercury, therefore we require that $a_\mathrm{ej}=0.3871$ au. If we permit 1\% variation in $a_\mathrm{ej}$ ($0.3867<a_\mathrm{ej}<0.3875$ au), we can calculate the ejecta velocity range satisfying such a condition. We generated $360\times180$ particle positions uniformly on the surface of Mercury's hill sphere with 360 longitudinal and 180 latitudinal positions. Approximately 50\% of ejection directions were feasible for both the perihelion and aphelion ejection.


Assuming the particles are leaving Mercury radially, we can calculate the velocity range that satisfies the 1\% range for $a_\mathrm{ej}$. We calculated the median $\delta v_\mathrm{ej} = 4.7$ m s$^{-1}$ for an impact occurring at Mercury's perihelion, and $\delta v_\mathrm{ej} = 5.9$ m s$^{-1}$ at Mercury's aphelion. Using Eq. \ref{EQ:Ejecta_Mass}, we get approximately 0.2\% for ejected material in such a velocity range; i.e., a factor of 500 material loss.


With a population of ejecta with orbits in the 1:1 MMR, we simulate their orbital evolution for 1000 years using 0.1-day time steps while keeping the same settings shown in Sec. \ref{SEC:Methods}. The number of particles still trapped in the 1:1 MMR after 100 years was approximately 0.1\%; i.e., 1 in 1000 was stable.

This brings us to an estimate that 1 in 1,000,000 particles ejected from Mercury's surface are transferred into quasi-stable 1:1 MMR with Mercury. In this exercise, we neglected the effects of radiation pressure, magnetic forces, and solar wind, which add more complexity to this largely unconstrained scenario for having assumed only the gravity of the Sun and 8 solar system planets. Moreover, due to lack of constraints we were forced to generalize many assumptions, such as the location and direction of the impact.

Using Eq. \ref{EQ:Ejecta_Mass} we estimated that an event resulting in a 40-km crater in diameter ejects $M_e = 5.60\times 10^{13}$ kg of material from which $M_\mathrm{MMR}=5.60\times10^7$ kg remains captured in the 1:1 MMR with Mercury. This value is $\sim4$ orders of magnitude smaller than the one we estimate from STEREO observations in Section \ref{SEC:Ring}. Looking for larger craters, we find that there are some fresh 100-km craters on Mercury's surface, such as Bartok and Amaral. Doubling the size of the source crater increases the ejected mass by a factor of $\sim10$.

Despite $3-4$ orders of magnitude discrepancy between our estimated mass captured in the 1:1 MMR with Mercury from a $40$-km crater impact event and the estimated mass of Mercury's dust ring, we showed that there is a theoretical possibility to transfer substantial amount of mass from Mercury's surface into the co-orbital resonance. Further developments in estimating impact yields for large craters on Mercury's surface, more complex simulations of the ejecta transfer into the co-orbital resonance, and additional observations of Mercury's dust ring might possibly narrow the gap between the model prediction and the current observation. In the next Section we provide the last missing puzzle piece and estimate the ages for the largest fresh craters currently known on Mercury's surface.

\section{Looking for youngest craters on Mercury}
\label{SEC:Craters_Ariel}
\citet{Kinczyk_etal_2020} previously classified all impact craters on Mercury with diameters $\ge40$ km on the basis of their degradation state. "Class 5" craters are interpreted to be the youngest craters because they have (i) bright ray systems, (ii) radially textured continuous ejecta blankets, (iii) fresh/crisp rim structures, central peak structures, and wall terraces, (iv) distinct floor-wall contacts, (v) floors that are the least partially covered with plains materials and contain hummocky deposits, (vi) well-defined continuous fields of crisp secondary craters, and (vii) no superposing craters at a pixel scale of $\sim166$ m \citep{Kinczyk_etal_2020}. Starting with the database of "Class 5" craters from \citet{Kinczyk_etal_2020}, we first down-selected our sample to include all Class 5 craters that have diameters $\gtrsim100$ km (Bart\'{o}k, Amaral, and Tyagaraja) to approach the order-of-magnitude mass of ejecta needed for the impact origin scenario described in the previous Section. We also selected craters that appeared particularly fresh on the basis of low maturity indices \citep{Neish_etal_2013} and/or low superposing crater densities (Basho, Debussy, and Unnamed A). The final sample of six analyzed craters in this article is listed in Table \ref{TABLE:Craters}. Note that Hokusai (98-km diameter) is not included in our analysis because optical maturity and S-band radar data \citep{Neish_etal_2013} \citep{Neish_etal_2013} suggest it is relatively older than both Debussy and Amaral, which are included.

\begin{deluxetable}{cccccccccc}




\tablecaption{Summary of the six craters analyzed in this work. Maximum estimated ages of the craters considered in this study derived from Poisson statistics and crater size-frequency distributions (CSFDs) using all identified superposing craters $\ge400$ m in diameter. Ages are considered to be maximums because they are derived from populations of small craters, maintaining the possibility that they are secondary craters. For each analyzed crater, we report the count surface area $A$, the number of superposing craters $\ge400$ m in diameter $N(\ge400\mathrm{~m})$, the number of superposing craters $\ge7$ km in diameter $N(\ge7\mathrm{~km})$ and the diameters of smallest $D_\mathrm{min}$ and largest $D_\mathrm{max}$ identified superposing crater diameters. \label{TABLE:Craters}}

\tablenum{2}

\tablehead{\colhead{Crater} & \colhead{Location} & \colhead{Diameter} & \colhead{Area} & \colhead{N} & \colhead{N} & \colhead{$D_\mathrm{min}$} & \colhead{$D_\mathrm{max}$} & \colhead{Poisson est. } & \colhead{CSFD est.}
\\
{} &
{($^\circ$N, $^\circ$E)} &
{(km)} &
{(km$^2$)} &
{($\ge400$ m)} &
{($\ge7$ km)} &
{(km)} &
{(km)} &
{age (Ma)} &
{age (Ma)} 
} 

\startdata
Bartók & [-29.26, -135.06] & 107.09 & 10950 & 28 & 1 & 0.55 & 15 & 233 $\pm$ 0.01 & 120 $\pm$ 23\\
Amaral & [-26.48, 117.90] & 101.32 & 4712 & 13 & 1 & 0.4 & 20 & 171 $\pm$ 0.02 & 54 $\pm$ 15\\
Tyagaraja & [3.90, -148.79] & 98.05 & 5400 & 59 & 0 & 0.4 & 1.5 & 263 $\pm$ 0.01 & 200 $\pm$ 25\\
Bashō & [-32.39, 189.54] & 75 & 2640 & 6 & 0 & 0.5 & 0.8 & 214 $\pm$ 0.03 & 34 $\pm$ 14\\
Debussy & [-33.95, 12.53] & 81 & 3740 & 4 & 0 & 0.75 & 1.3 & 259 $\pm$ 0.04 & 102 $\pm$ 50\\
Unnamed A & [0.36, 17.58] & 40.06 & 1074 & 9 & 0 & 0.4 & 0.9 & 246 $\pm$ 0.02 & 75 $\pm$ 25\\
\enddata




\end{deluxetable}

Because they formed during the Kuiperian era, all Class 5 craters have an upper-limit age of 100-300 Myr \citep{Banks_etal_2017}. Here, we provide more accurate estimates of the ages of the six dust-ring-forming candidates using independent techniques: Poisson statistics and crater size-frequency distributions (CSFDs). Using monochrome images acquired by the Mercury Dual Imaging System (MDIS) narrow-angle camera (NAC) \citep{Hawkins_etal_2007}, we surveyed  the craters of interest from Table \ref{TABLE:Craters}, and identified all superposing craters with diameters $\ge 400$ m that can confidently be resolved in the surveyed NAC images (see Table \ref{TABLE:Crater_Images}). We used Poisson statistics and Bayesian inference to describe the relative likelihoods of possible ages of the crater surfaces \citep{Michael_etal_2016}. The likelihood function determines the time-resolved probability of a given crater observation within the chronology model for the set of observed craters $\mathcal{D}$, as divided into $n$ bins for any given time $t$, and is expressed by \citet{Michael_etal_2016} in their Eq. (8):
\begin{equation}
    \mathrm{pr}(\mathcal{D},t) \propto \mathrm{exp}(-A[C(D,t)]^{D_\mathrm{min}}_{D_\mathrm{max}} \{C(D=1,t)\}^{n_\mathrm{\mathcal{D}}},
\end{equation}
where $A$ is the crater accumulation area, $C(D,t)$ is the cumulative form of the production function, and $D$ is the crater diameter. We adopt the production function presented by \citet{LeFeuvre_Wieczorek_2011} and use 10,000-year time steps. The spreads of possible ages and their relative likelihoods are equally valid for surfaces that have no craters, only a few craters, or many craters, and uncertainties stem from the predictions of the chronology model itself \citep{Michael_etal_2016}. Uncertainties are reported as $1-\sigma$ of the median to include 68.13\% of the function describing the likelihood that the surface has a particular age.

We also used CSFDs to estimate the crater formation ages. Crater counting was performed using the NAC images listed in Table \ref{TABLE:Crater_Images} in the JMARS software \citep{Edwards_etal_2011}. Model ages were then calculated using the CraterStatsII program \citep{Michael_Neukum_2010} with the \citet{LeFeuvre_Wieczorek_2011} chronology system and 400-m minimum crater diameters. Absolute model ages were estimated using a cumulative fit to the counted craters from pseudo-log binning, employing a cumulative resurfacing condition and porous scaling. In the CraterStatsII program, ages are estimated by dividing the crater population into discrete diameter intervals, plotting the crater density for each interval, and determining a best-fit model isochron \citep{Michael_Neukum_2010}. We report the crater ages with $\operatorname{1-\sigma}$ uncertainties, which are calculated as $1/n^{0.5}$, where $n$ is the number of craters in a given diameter interval used for the age fit. Uncertainties are then translated into an error in the age with respect to the chronology function. These uncertainties are derived from counting statistics alone and so do not incorporate systematic errors associated with the chronology function. Because the number of the observed craters is relatively low, the reported uncertainty from crater counting is inherently large. Thus, a major advantage of utilizing Poisson statistics and Bayesian inference is that an exact evaluation of the crater chronology model can be made and no minimum crater counts are required \citep{Michael_etal_2016}.

We note that low spatial resolution and non-ideal illumination conditions of surveyed images can lead to biases of cratering observations that are used as inputs in the likelihood model \citep{Williams_etal_2018}. To mitigate this, we surveyed each crater under a range of incidence and illumination angles. The presence of secondary craters can also bias our observations, and secondary craters on Mercury can be difficult to distinguish because they can be several kilometers in diameter, and appear more circular in shape and more isolated in distribution than secondary craters on other terrestrial bodies \citep{Strom_etal_2008, Strom_etal_2011,Xiao_etal_2014,Xiao_2016}. Although we excluded asymmetric craters and obvious secondary impact clusters, chains, and rays, it is still very possible that some background secondaries are included in our cratering observations. On Mercury, secondary craters become increasingly more abundant at diameters $<10$ km and are particularly prevalent in crater populations at diameters $<7$ km \citep{Strom_etal_2008}. We only identified 2 superposing craters $>7$ km (1 in Bart\'{o}k and 1 in Amaral), making it possible that any of the other identified craters could be secondaries as well. Including secondary craters in our model would skew the derived ages to be older than their real age. Furthermore, relatively young surfaces (like those considered here) are expected to have the highest uncertainties associated with secondary cratering \citep{Hartmann_Daubar_2017}. Thus, given the small sizes of the identified superposing craters, we consider the estimated ages derived here to be the maximum crater ages.

\subsection{Estimated crater ages}
Ages derived from Poisson statistics suggest that all of the analyzed craters are most likely to have formed between $\sim170$ Ma and $\sim260$ Ma, with Amaral crater forming most recently $\mathcal{T}_\mathrm{Pois} = 171\pm0.02$ Ma. This is consistent with analysis of optical maturity properties that also suggests Amaral is among the youngest large craters on Mercury \citep{Neish_etal_2013}. Recall that the crater ages derived here are considered to be the maximum age estimates because the majority of superposing craters identified in our analysis are smaller than 7 km in diameter and therefore may be secondary impacts.

When considering the CSFD ages, Amaral appears to be the second youngest crater in our sample ($\mathcal{T}_\mathrm{CSFD} = 54\pm15$ Ma) and is surpassed by Bashō with $\mathcal{T}_\mathrm{CSFD} = 34\pm14$.  While these younger ages of craters look promising for our recent impact scenario described in Sec. \ref{SEC:Dust_From_1:1} that requires an impact within the last 15 Myr, the ages derived from CSFDs and Poission statistics do not overlap, even considering the larger uncertainties associated with CSFD fitting. Due to the apparent advantages of Poisson statistics over CSFD fitting described in the previous Section, we feel that more accurate constraints on crater ages are required to confidently reject the recent impact scenario for the origin of Mercury's dust ring. With the BepiColombo spacecraft on its was to Mercury, we can expect new exciting data sets in several years that may aid in further constraining crater ages \citep{Rothery_etal_2020}.

\section{Can exoplanetary systems exhibit similar signatures to Mercury's circumsolar dust ring?}
The possibile  occurrence of an impact-generated circumsolar dust ring connected to a low-mass, eccentric planet orbiting close to the host star raises interesting consequences for exosolar planetary systems at different stages of evolution. The solar system is a very mature planetary system and the impact rates on terrestrial planets are orders of magnitude smaller than their exoplanetary analogs with debris signatures observable today \citep{Hughes_etal_2018}. In Fig. \ref{FIG:Residence_Time} we showed that the meteoroid stability in the 1:1 MMR with Mercury can reach 15 Myr, and from both numerical experiments \citep[e.g.,][for particles in 1:1 MMR with Venus]{Pokorny_Kuchner_2019, Pokorny_Kuchner_Sheppard_2020} and analytical predictions \citep{Dermott_etal_1994} we can extrapolate that the stability increases with the planet's distance from the host star, the planet's own mass, and with smaller planetary eccentricity. Therefore, if such impacts occur on exoplanets embedded in orders of magnitude more impactors, then the resulting circumsolar ring might be getting stronger in time until the particle loss from radiation effects and collisions equals the transfer rate to the co-orbital resonance.

However, this is only valid for planets smaller than or comparable in size to Earth and close to the host star. If the planet is too massive, the higher escape velocity will prevent any ejecta from leaving the planet's gravity well. If the planet is too far from its host star, the impactors will not likely have impact velocities high enough for material to transfer from the surface to the co-orbital resonance. The existence of a non-tenuous atmosphere would further prevent the ejecta escape from the planetary surface. Therefore, until we have a more complete inventory of exoplanets and debris disks, we can only speculate about the occurence of Mercury's dust ring analogs.

\section{Discussion}
\subsection{Potential future missions observing Mercury's dust ring}
Additional observations of Mercury's dust ring would provide important new data required for more comprehensive modeling and model fitting. Currently, there are several space missions capable of detecting Mercury's dust ring: (a) the Parker Solar Probe has completed its first 10 close encounters with the Sun \citep{Fox_etal_2016} where each encounter passes through Mercury's orbit potentially giving an opportunity to observe Mercury's dust ring both from the outside and inside similarly to the all-sky observation of Venus's dust ring \citep{Stenborg_etal_2021_Venus_Ring}; (b) the Solar Orbiter mission \citep{Muller_etal_2020} - thanks to its unique, inclined orbit - might be able to observe Mercury's dust ring from a location away from the ecliptic and produce crucial data for characterization of Mercury's dust ring; and finally (c) BepiColombo will have a unique opportunity to detect Mercury's dust ring via remote sensing as well as in-situ with an onboard dust detector Mercury Dust Detector (MDM) \citep{Kobayashi_etal_2020} when it approaches and orbits the planet.

\subsection{Dust ring stability with respect to collisions with Zodiacal Cloud particles}
As we showed in Sec. \ref{SEC:Residence_Time}, particles trapped in the 1:1 MMR with Mercury have a maximum residence time of $T_\mathrm{res}\sim15$ Myr. All these trapped particles are continuously sweeping through the inner sections of the ZC and thus have the potential to be collisionally fragmented or destroyed before they become dynamically unstable. Using traditional methods to estimate collisional lifetimes of particles sharing Mercury's orbit \citep{Grun_etal_1985, Steel_Elford_1986}, we estimate the collisional lifetime of particles to be $\sim100\times$ shorter than the dynamical lifetimes shown in Fig. \ref{FIG:Residence_Time}. Should these collisional lifetimes be valid, the age of Mercury's dust ring would be $<1$ Myr and thus very improbable based on our impact origin hypothesis. Fortunately, many recent dynamical model works have shown that the collisional lifetimes in the inner solar system for asteroidal and cometary meteoroids should be $20-100\times$ longer \citep[e.g.,][]{Nesvorny_etal_2011JFC,Pokorny_etal_2014, Soja_etal_2019} and based on our estimates the collisional lifetimes of meteoroids in the co-orbital resonance with Mercury are comparable to their residence times (Sec. \ref{SEC:Residence_Time}).
For these reasons, we think that collisions of particles locked in 1:1 MMR with Mercury with inner ZC do not prevent the existence of Mercury's dust ring based on our impact origin hypothesis. ZC particles will erode and disrupt a certain portion of Mercury's dust ring and diminish its magnitude in time in timescales similar to dynamical transport of particles from the co-orbital resonance with Mercury.

\subsection{Other potential craters connected to Mercury's dust ring}
Starting with the crater database from \citet{Kinczyk_etal_2020} we provided age estimates for six "Class 5" (fresh-looking) craters with $D>40$ km. Many other fresh-looking craters smaller than 40 km are also present on the surface of Mercury. As described by the crater production function \citep[e.g.,][]{LeFeuvre_Wieczorek_2011}, the number of small craters that formed in Mercury's recent past is larger than the number of large craters, and therefore some of these smaller craters would have formation ages that are younger than those derived for the larger craters summarized in Table \ref{TABLE:Craters}. While a single one of these craters cannot provide enough ejecta to form the currently observed dust ring in one event, it is possible that multiple smaller impacts close enough in time ($\sim10$ Myr) may have contributed material that collectively amounts to the currently observed dust ring.

\section{Conclusions}
In this work we test several origin hypotheses for the existence of Mercury's circumsolar dust ring. We find that the following origin scenarios do not have the ability to create a circumsolar dust structure close to Mercury: a) sporadic meteoroid background (Sec. \ref{SEC:SPORADIC}); b) recent asteroidal and cometary activity (Sec. \ref{SEC:RECENT_ACTIVITY}; and c) hypothetical dust sources originating in MMRs outside Mercury (Sec. \ref{SEC:MMRs}). These three origin scenarios include all conceivable currently known sources of dust particles beyond Mercury's orbit.

We find that the only source population able to create the circumsolar dust ring with its observed brightness and shape is a population of dust particles and meteoroids co-orbiting with Mercury (Sec. \ref{SEC:Dust_From_1:1}). Our model agrees well with the STEREO observations from \citet{Stenborg_etal_2018} and suggests that the dust ring extends beyond solar elongations $23^\circ$, which was the limit of STEREO observations (Sec. \ref{SEC:Model_Comparison_To_Stereo}). Our model also suggests that the 2P/Encke meteoroid stream is superimposed on Mercury's dust ring and the magnitude of this stream with respect to the dust ring itself could be inferred from observations at larger solar elongations (Sec. \ref{SEC:2PEncke}).

We find that any primordial population of larger bodies that could source the observed dust is not dynamically stable, where our calculations suggest that dynamical stability of any small body in the co-orbital resonance with Mercury is $<20$ Myr (Sec. \ref{SEC:Residence_Time}). Therefore, we need an alternative and recent dust-generating source with a potential to enter the co-orbital resonance with Mercury. Our calculations suggest that a recent impact on Mercury has a non-zero probability of transporting Hermean ejecta into the 1:1 MMR; however, the transport efficiency and the mass of ejecta derived from the sizes of recent craters on Mercury's surface give a few orders of magnitude less material than needed to explain the current observed density of the dust ring (Sec. \ref{SEC:RecentImpact}).

We analyze the surface of Mercury and estimate the ages of the six candidate craters that are larger than 40 km in diameter and appear to be geologically fresh (Sec. \ref{SEC:Craters_Ariel}). Using the crater size-frequency distribution age estimator we find two craters with estimated ages younger than 50 million years inside 1-$\sigma$ uncertainty range: Amaral with an estimated age $54\pm15$ Ma and a diameter of 101.32 km and Bash\={o} with an estimated age $34\pm14$ Ma and a diameter of 75 km. Using an alternative age estimator (Poission statistics) pushes the crater ages beyond 150 Ma, making their connection with Mercury's dust ring improbable.

We expect that Mercury's dust ring is not an isolated phenomenon but might be a common event in many mature exoplanetary systems where airless planets are bombarded by kilometer-sized impactors at impact velocities exceeding the planet's escape velocities. In our solar system, Venus' and Earth's atmospheres preclude the impact-driven creation of circumsolar dust rings, which leaves Mercury as the only candidate for this interesting phenomenon.

\software{
\texttt{gnuplot} (\url{http://www.gnuplot.info}) $\bullet$
\texttt{swift} \citep{Levison_Duncan_2013} $\bullet$ \texttt{matplotlib} (\url{https://www.matplotlib.org}) \citep{Hunter_2007} $\bullet$
\texttt{SciPy} (\url{https://www.scipy.org})
}

\facilities{NASA Center for Climate Simulation (NCCS), NASA Advanced Data Analytics PlaTform (ADAPT)}

\begin{acknowledgments}
\noindent Funding: P.P. and M.J.K. were supported by NASA ISFM EIMM award, the NASA Cooperative Agreement 80GSFC21M0002 and NASA Solar System Workings award No. 80NSSC21K0153. A.N.D was supported by an appointment to the NASA Postdoctoral Program at Ames Research Center, administered by Oak Ridge Associated Universities under contract with NASA.
\end{acknowledgments}

\appendix
\section{List of NAC images used for crater identification}
\begin{deluxetable}{cccccccccc}




\tablecaption{List of images used for crater identification. \label{TABLE:Crater_Images}}

\tablenum{A1}

\tablehead{\colhead{Crater} & \colhead{Location} & \colhead{Image ID} & \colhead{Map scale}
\\
{} &
{($^\circ$N, $^\circ$E)} &
{} &
{(m)} &
} 
\startdata
Crater & Location (°N, °E) & Image ID & Map scale (m) \\
\hline
Barlok & -29.26, -135.06 & EN0257675926M & 127 \\
 &  & EN0244519749M & 120.2 \\
 &  & EN0229323911M & 117.9 \\
 &  & EN0244490765M & 115.1 \\
 &  & EN1024444389M & 118.9 \\
 &  & EN1024415895M & 127.4 \\ \hline
Amaral & -26.48, 117.90 & EN0251575891M & 79.8 \\
 &  & EN0236873588M & 79.5 \\
 &  & EN0236831211M & 79.5 \\
 &  & EN0236831181M & 77.4 \\\hline
Tyagaraja & 3.90, -148.79 & EN1037287658M & 31 \\
 &  & EN1037287654M & 31.4 \\
 &  & EN1037316487M & 30.9 \\
 &  & EN1037316503M & 30.2 \\
 &  & EN1037287662M & 30.6 \\
 &  & EN1037316455M & 32.1 \\\hline
Bashō & -32.39, 189.54 & EN0258596921M & 155.2 \\
 &  & EN0258625867M & 153.2 \\
 &  & EN0227934756M & 169.5 \\
 &  & EN0242918833M & 234.8 \\\hline
Debussy & -33.95, 12.53 & EN1034606672M & 99.5 \\
 &  & EN1034924248M & 74.6 \\
 &  & EN1034664579M & 88.8 \\
 &  & EN1034606722M & 97.5 \\
 &  & EN1034606697M & 98.5 \\
 &  & EN1017272071M & 127.4 \\\hline
Unnamed A & 0.36, 17.58 & EN1014789424M & 46.2 \\
 &  & EN1014817948M & 55.6 \\
 &  & EN1014818239M & 46.7 \\
 &  & EN1014903758M & 82.8 \\
 &  & EN1014904287M & 61.9 \\
 &  & EN1014904312M & 61 \\
 &  & EN1014904632M & 50.8 \\
 &  & EN1014789138M & 54.8 \\
 &  & EN1014760387M & 52.4 \\
 &  & EN1014702801M & 50.1 \\
 &  & EN1014674229M & 43.2 \\
 &  & EN1014673973M & 50.6 \\
 &  & EN1030027775M & 47.3 \\
\enddata




\end{deluxetable}

\end{document}